\documentclass[utf8]{FrontiersinHarvard} 

\usepackage{url,hyperref,lineno,microtype,subcaption}
\usepackage[onehalfspacing]{setspace}
\usepackage{soul}

\usepackage[thicklines]{cancel}

\def\keyFont{\fontsize{8}{11}\helveticabold }
\def\firstAuthorLast{Zhao {et~al.}} 
\def\Authors{Huilin Zhao\,, Sungchil Yang\,$^{*}$ and Chi Chung Alan Fung\,$^{*}$}
\def\Address{ Department of Neuroscience, City University of Hong Kong, Tat Chee Avenue, Kowloon, Hong Kong, China }
\def\corrAuthor{Corresponding Author}

\def\corrEmail{sungchil.yang@cityu.edu.hk, alan.fung@cityu.edu.hk}

\begin{document}
\onecolumn
\firstpage{1}

\title {Short-Term Postsynaptic Plasticity Facilitates Predictive Tracking in Continuous Attractors}

\author[\firstAuthorLast ]{\Authors} 
\address{} 
\correspondance{} 

\extraAuth{}

\maketitle

\begin{abstract}

The N-methyl-D-aspartate receptor (NMDAR) is a crucial component of synaptic transmission, and its dysfunction is implicated in many neurological diseases and psychiatric conditions. NMDAR-based short-term postsynaptic plasticity (STPP) is a newly discovered postsynaptic response facilitation mechanism. Our group has suggested that long-lasting glutamate binding of NMDAR allows input information to be held for up to 500 ms or longer in brain slices, which contributes to response facilitation. However, the implications of STPP in the dynamics of neuronal populations remain unknown. In this study, we implemented STPP in a continuous attractor neural network (CANN) model to describe the neural information encoded in neuronal populations. Unlike short-term facilitation, which is a kind of presynaptic plasticity, the temporally enhanced synaptic efficacy induced by STPP destabilizes the network state of the CANN by increasing the mobility of the system. This nontrivial dynamical effect enables a CANN with STPP to track a moving stimulus predictively, i.e., the network state responds to the anticipated stimulus. Our findings reveal a novel STPP-based mechanism for sensory prediction that can help develop brain-inspired computational algorithms for prediction.

\tiny
 \keyFont{ \section{Keywords:} Computational Models, Synaptic Plasticity, Attractor Models, Predictive Coding, NMDA Receptors} 
\end{abstract}

\section{Introduction}
{The N-methyl-D-aspartate receptor (NMDAR) is a Ca$^{2+ }$-permeable, ligand-gated ion channel found mainly in the postsynaptic membrane of neurons, facilitating synaptic transmission \citep{Hunt2012, Paoletti2013, Yao2022}. It consists of two GluN1s and two GluN2s (GluN2A-D) and becomes active when two glutamates simultaneously bind to two GluN2s \citep{Monyer1992, Paoletti2013, VYKLICKY2014}. When activated, it produces a regenerative Ca$^{2+ }$ spike, which contributes to signal integration occurring in postsynaptic dendrites \citep{Schiller2000, Branco2011, Yang2015, Noh2019}. The heterogeneity in NMDAR subunits generates the diversity of its properties and functions \citep{Paoletti2013}. NMDARs are crucial for neurotransmission and neuronal communication in the nervous system \citep{Paoletti2007, Hansen2018}. Their} dysfunction resulting from hyperactivity, hypofunction, abnormal subunit expression, altered receptor trafficking or localization may contribute to a variety of neurological diseases and psychiatric conditions \citep{Zhou2013}, such as Huntington's disease \citep{Burtscher2021}, Alzheimer’s disease \citep{Liu2019}, depression \citep{Marsden2011, Adell2020}, schizophrenia \citep{Lisman2008, Nakazawa2020, Adell2020}, and ischemic stroke \citep{Chen2008}. NMDARs also play an essential role in synaptic plasticity, which refers to the strengthening or weakening of electrical postsynaptic responses over time in response to past synaptic activities \citep{Citri2007}. Furthermore, NDMARs profoundly influence synaptic functions that underlie high-level cognitive functions \citep{Paoletti2013, Bertocchi2021}. For example, selective modulation of subunits of NMDARs impaired long-term potentiation, a typical type of long-term synaptic plasticity, in striatal synapses and thus caused a visuospatial learning deficit \citep{Durante2019}. Another example is the repetitive-training enhanced NMDAR-mediated synaptic transmission in the medial prefrontal cortex that was involved in social memory retrieval \citep{Zhang2022}. 

\begin{figure}[h!]
        \begin{center}
                \includegraphics[width=17.5cm]{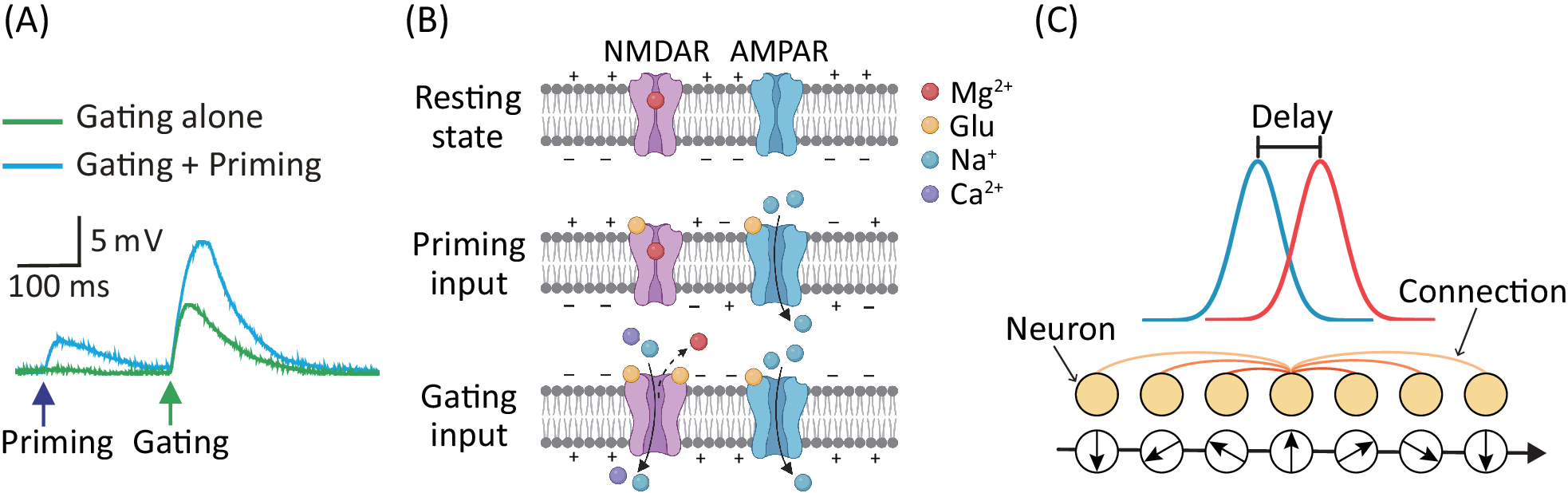}
        \end{center}
        \caption{(A) Evidence of higher postsynaptic response when priming plus gating (cyan trace) glutamate uncaging is applied than when gating alone (green trace) is applied was observed in a hippocampus dendrite \citep{Yang2016}. (B) Schematics of the mechanism of STPP, which is used to interpret the observed increase in postsynaptic response through the actions of receptors on the postsynaptic membrane \citep{Yang2018}. Top: NMDARs and AMPARs are closed in the resting state. Middle: AMPARs are then activated by the priming input, whereas the NMDARs do not open because they are blocked by magnesium ions. However, NMDARs enter the bound-but-blocked state in which the priming input information can be stored for up to 500 ms or longer. Bottom: When a gating input arrives, the postsynaptic membrane is further depolarized and magnesium is removed, shifting the NMDARs from the bound-but-blocked state to the open state. (C) Illustration of a CANN that models neural activities involved in continuous stimuli, e.g., head direction. Neurons are arranged in the network in accordance with their preferred stimuli. The excitatory connections among neurons are translationally invariant in the space of stimulus values. The CANN is able to track moving stimuli, i.e., the red bump-shaped curve, but the neuronal activities i.e., the blue bump-shaped curve, always lag behind the stimuli due to the neural response delay \citep{Wu2005}.}\label{fig:fig1}
\end{figure}

Although it is widely accepted that NMDAR-mediated Ca$^{2+ }$ signaling contributes to long-term synaptic plasticity in postsynaptic neurons \citep{Hunt2012, Granger2014, Volianskis2015}, its effects on short-term synaptic plasticity are gradually being uncovered and studied \citep{Santos2012, Yang2014, Yang2016}. Typically, short-term synaptic plasticity is attributed to the difference in the time constants between neuronal signaling and recovery of neurotransmitter availability \citep{Zucker2002, Mongillo2008}. Neurotransmitter release from presynaptic neurons is primarily responsible for this neurobiological mechanism. {Although neuronal plasticity is observed by presynaptic factors for neurotransmitter release, receptors directly modulating the efficacy of postsynaptic currents are situated on the postsynaptic side, which is the postsynaptic NMDARs. According to our previous studies \citep{Yang2014,Yang2016,Yang2018}, NMDAR-dependent short-term postsynaptic plasticity (STPP) is proposed to serve as a neurobiological mechanism for signal amplification, particularly in linearly connected circuits such as the hippocampus and cortices. Such signal amplification can be timely achieved through STPP, which enables the faithful transmission of extrinsic information-bearing sensory inputs and the integration of extrinsic sensory inputs (i.e., a priming input) with intrinsic activity (e.g., a brain rhythm or gating input). Unlike the presynaptic mechanism underlying a feedback circuit for signal amplification, this STPP executes a feedforward process to carry out an efficient and timely signal amplification and cascade.}
One of the observed pieces of evidence showing STPP in a hippocampus dendrite \citep{Yang2016} is shown in Figure \ref{fig:fig1}A. The postsynaptic response was higher when a second (gating input) glutamate uncaging followed the prior (priming input) uncaging (cyan trace) than when a gating input alone (green trace) was applied. This enhancing effect was eliminated by Ifenprodil, a blocker of GluN2B \citep{Yang2016}. An underlying mechanism (Figure \ref{fig:fig1}B) for short-term signal amplification has been proposed in our earlier publication \citep{Yang2018}. The $\alpha$-amino-3-hydroxy-5-methyl-4-isoxazolepropionic acid receptor (AMPAR) is another kind of glutamate receptor on the postsynaptic membrane \citep{Diering2018}. NMDARs and AMPARs are closed because the membrane is in the resting state. The AMPARs are then activated by a priming input, i.e., the first glutamate release from the presynaptic neurons, whereas the NMDARs do not open because they are blocked by magnesium ions. However, the input information can be stored via glutamate binding, and the NMDARs enter the bound-but-blocked state for up to 500 ms (or longer) \citep{Yang2018, Yang2014}. Subsequently, the gating input, i.e., the second glutamate release, is strong enough to further depolarize the postsynaptic membrane, and thus magnesium is removed. Simultaneous membrane depolarization and glutamate binding shift the NMDARs from the bound-but-blocked state to the open state. That is, NMDARs are more likely to open with the previously stored priming input plus gating input than with the gating input alone. The NMDAR-mediated Ca$^{2+}$ current, which is much stronger than the AMPAR current, is thought to contribute to the response enhancement. {Additionally, when the priming input was so large to induce NMDAR-mediated Ca$^{2+}$ current, there was no significant enhancement of the second response under the same level of gating input (Figure 1B in \citep{Yang2016}). 
Therefore, STPP depends on the state of postsynaptic NMDARs and the evocation history of postsynaptic currents, but not necessarily the firing history of a {\it particular} presynaptic neuron. In this regard, STPP has been suggested to enable signal amplification and integration in synaptic transmission, but the role of STPP in the dynamics of neuronal populations remains unknown.}

We aim to investigate the possible effects of STPP in neuronal populations on neural information processing. To this end, we introduce the continuous attractor neural network (CANN) as our model for neural information representation \citep{Wu2005, Wu2016}. The CANN is capable of extracting the information encoded by a population of neurons, which allows neural information processing to be analyzed and simulated computationally \citep{Deneve1999, Wu2005}. This model has been successfully used to depict the encoding of continuous stimuli in neural systems, such as movement direction \citep{Georgopoulos1993}, head direction \citep{BenYishai1995, Zhang1996, Stringer2002a}, the spatial location of objects \citep{Samsonovich1997, Stringer2002b} and spatial integrated information including location, direction and distance \citep{Burak2009}. A CANN is shown in Figure \ref{fig:fig1}C. In the network, a bump-shaped tuning curve (blue curve) represents the neural activities of the population of neurons in response to the external stimuli (red curve) at a given time. Because different neurons can be characterized by their preferred stimuli, e.g., different head directions represent different head-direction (HD) cells, the response curve is a function of the preferred stimuli of the neurons in a population. Among neurons, there are excitatory connections that are translationally invariant in the space of stimulus values. These translation-invariant connections enable a CANN to hold a continuous family of stationary states (attractors). The stationary states of the neural system form a continuous parameter space in which the system is neutrally stable. This property allows the neutral system to track time-varying stimuli smoothly \citep{Wu2008, Fung2010}. However, the tracking always lags behind the stimulus due to the time needed for neuronal responses and neuronal interactions \citep{Wu2005, Wu2016}. Because CANNs can track moving objects smoothly, they have been used to shed light on the potential neural correlates of effective tracking of moving objects \citep{Zhang1996, Zhang2012, Fung2012, Mi2014, SFard2015}. For instance, inhibitory feedback modulations such as short-term depression (STD) of neuronal synapses \citep{Fung2012}, spike frequency adaptation (SFA) in neural firing \citep{Mi2014}, and negative feedback from a connected network \citep{Zhang2012} have enabled predictive tracking of sensory input with CANNs. A path integration mechanism combined with CANNs has predicted future movement locations \citep{SFard2015}. These neural predictions, i.e., anticipations, of continuously moving objects are powerful strategies for time delay compensation \citep{Nijhawan1994, Bassett2005, Sommer2006} and thus maintain effective and efficient perceptual functions and visual/motor control. 

In this study, we applied the STPP mechanism to CANNs and examined the tracking dynamics when they tracked moving stimuli. The stimuli could be head directions, object locations or navigation information such as speeds and directions of movements. To simplify our study, we built a one-dimensional CANN model and used head directions as our representative stimuli. Through the simulation experiments and analysis, {we found that STPP-induced enhancing effect on neurons around the hillside of synaptic input profile enables CANNs to anticipate the movements of moving stimuli.} Unlike the mechanisms for anticipatory tracking based on inhibitory feedback modulations \citep{Zhang2012, Fung2012, Mi2014}, STPP is a feedforward modulation driven by the inherent features of NMDARs. Our findings suggest a novel mechanism for anticipatory tracking. The reliable signal transmission enabled by STPP provides a new framework that has the potential to help conceptualize a network mechanism for sensory prediction and develop brain-inspired computational algorithms for prediction.

\section{Model and Simulation}
\subsection{The Model}
In our study, we used CANNs \citep{Wu2005, Fung2010} to investigate the influence of STPP \citep{Yang2016} on the dynamics of neuronal populations.
{In this model, the dynamics of synaptic input $u\left(x,t\right)$ of neurons with preferred stimuli $x$ at time $t$ is defined as}
\begin{equation}
\tau_\text{s}\frac{\partial u\left(x,t\right)}{\partial t} =-u\left(x,t\right)+\left[1+S(x,t)\right]I^{\text{tot}}\left(x,t\right), \label{eq:dudt}    
\end{equation}
where $\tau_\text{s} \approx 10\text{ ms}$ is the neuronal time constant \citep{Koch1996}. 
{ $\left[1+S(x,t)\right]$ models the efficacy of presynaptic neurons in the postsynaptic input evocations.}
$I^{\text{tot}}\left(x,t\right)$, the total input to the neurons from the external input and lateral interactions within the neuronal system, is given by
\begin{equation}
I^{\text{tot}}\left(x,t\right) =\int dx^{\prime}J\left(x,x^{\prime}\right)r\left(x^{\prime},t\right)+I^{\text{ext}}\left(x,t\right).\label{eq:Itot}
\end{equation}
$I^{\text{ext}}\left(x,t\right)$ is the external input, which is defined in the later subsection. 
$J\left(x,x^{\prime}\right)$, the excitatory connection between neurons at $x$ and $x^\prime$, is given by
\begin{equation}
J\left(x,x^{\prime}\right)=\frac{1}{\sqrt{2\pi}a}\exp\left(-\frac{\text{dist}\left(x,x^\prime\right)^{2}}{2a^{2}}\right),
\label{eq:Jxx}
\end{equation}
where $\text{dist}(x,x^\prime)$ describes the distance between $x$ and $x^\prime$ depending on the characteristics of the stimulus, which is defined in the following subsection. $a=0.5$ represents the range of the connection in the preferred stimuli space \citep{Fung2010} and also controls the tuning width of the attractor states. $r\left(x,t\right)$ is the neuronal response of neurons with preferred stimuli $x$ at time $t$. It also encodes the firing rate and is defined as a function of $u\left(x,t\right)$:
\begin{equation}
r\left(x,t\right)=\Theta\left[u\left(x,t\right)\right]\frac{u\left(x,t\right)^{2}}{1+\frac{1}{8\sqrt{2\pi}a}k\int dx^{\prime}u\left(x^{\prime},t\right)^{2}},
\label{eq:r}
\end{equation}
where $k$ controls the divisive inhibition modeled in the denominator of $r\left(x,t\right)$ \citep{Deneve1999, Wu2005}. $\Theta$ is the step function.
{ One should note that $I^{\text{tot}}\left(x,t\right)$ is a total input integrating the contributions of excitatory and inhibitory signals from a population of neurons regardless of the type of receptors. Also, since the inhibition is divisive, $\left[1+S\left(x,t\right)\right]$ in Equation \eqref{eq:dudt} modulates the excitatory input, which is consistent with the experimental results by Yang et al. \citep{Yang2016}.}

In Equation \eqref{eq:dudt},
{$S\left(x,t\right)$ is the enhancing modulation that abstractly models the effect on the synaptic input due to the opening of NMDARs from the bound-but-blocked state. It represents the temporal enhancement due to STPP. Therefore, $\left[1+S\left(x,t\right)\right]I^{\text{tot}}\left(x,t\right)$ models the synaptic input evocation triggered by presynaptic neuronal activity with an enhancement due to STPP \citep{Yang2016}. This term will be further discussed in the Discussion section. The corresponding latent modulation $Q\left(x,t\right)$ of $S\left(x,t\right)$ abstractly models the proportion of NMDARs that enter the bound-but-blocked state. The dynamics of $S\left(x,t\right)$ and $Q\left(x,t\right)$ are defined by}{
\begin{align}
\frac{\partial S\left(x,t\right)}{\partial t} & =-\frac{S\left(x,t\right)}{\tau_{1}}+\alpha Q\left(x,t\right)f_{S}\left[r\left(x,t\right)\right],\label{eq:dSdt}\\
\frac{\partial Q\left(x,t\right)}{\partial t} & =-\frac{Q\left(x,t\right)}{\tau_{2}}-\alpha Q\left(x,t\right)f_{S}\left[r\left(x,t\right)\right]+\beta\left[1-Q\left(x,t\right)\right]f_{Q}\left[I^{\text{tot}}\left(x,t\right)\right],\label{eq:dQdt}
\end{align}
}
where $\tau_1 = 50\text{ ms}$ is the time constant of NMDAR \citep{Hestrin1990},{ which controls the effective duration of the enhancing effect on postsynaptic input.} $\tau_2 = 500\text{ ms}$ is the time constant of the latent modulation of STPP \citep{Yang2018}, which controls its effective duration. The parameters $\alpha$ and $\beta$, which control the rates of $S\left(x,t\right)$ and $Q\left(x,t\right)$, respectively, are critical for adjusting the effects of STPP. From a physiological point of view, $\alpha$ can be considered the average opening rate of NMDARs and {$\beta$ is the average transition rate of NMDARs from the glutamate-unbound state to the bound-but-blocked state, which is determined by the speed and efficiency of glutamates to bind to NMDARs.}
$f_{S}$ and $f_{Q}$ are activation functions of $S\left(x,t\right)$ and $Q\left(x,t\right)$, respectively. $f_S$ defines the raise of the enhancing modulation $S\left(x,t\right)$. {We designed its form based on two considerations: (1) in accordance with the STPP mechanism, the removal of magnesium, which depends on the membrane potential, enhances the excitatory postsynaptic potential by opening NMDARs \citep{Jahr1990, VargasCaballero2004}. Hence, $f_S$ should be sigmoid-shaped based on membrane potential, modeling the magnesium removal efficacy of the postsynaptic membrane. (2) However, in the CANN, the membrane potential is implicit. Based on the study conducted by Latimer's group \citep{Latimer2019}, the average membrane potential can be approximated by the firing rate because they share a rectified-linear relation. In CANNs, $r\left(x,t\right)$ represents the neuronal activity, which corresponds to the firing rate of spiking neurons. In contrast, the synaptic input $u\left(x,t\right)$ integrates neuronal activities from lateral neurons and external input, which has a non-linear relation with the firing rate in the presence of divisive inhibition. Therefore, we chose $r\left(x,t\right)$ as a proxy to represent the average membrane potential for a population of neurons sharing the same preferred stimulus $x$. As a consequence, $f_S$ is defined to be a cumulative distribution function of $r\left(x,t\right)$:
\begin{equation}
f_S\left[r\left(x,t\right)\right]=\frac{1}{\sqrt{2\pi } } \int_{-\infty }^{\frac{r\left(x,t\right)-r_0}{\sigma_S} } dt^{\prime}\exp\left(-\frac{{t^{\prime}}^2}{2}\right),
\label{eq:fS}
\end{equation}}
where $r_0$ and $\sigma_S$ are the mode and scale of its probability density distribution, respectively. $f_Q$ is a function that drives the latent modulation $Q\left(x,t\right)$ depending on the total synaptic input.
{Given that insufficient input prevents NMDARs from entering the bound-but-blocked state and too much input results in strong depolarization and subsequent opening of NMDARs, moderate input can drive as many NMDARs as possible to enter the bound-but-blocked state. Nevertheless, strong input would have a chance to leave a portion of NMDARs in the bound-but-blocked state. Therefore, we considered a log-normal distribution with a right skew to strong input to describe the relationship between the probability of entering the bound-but-blocked state and total synaptic input $I^{\text{tot}}\left(x,t\right)$:
\begin{equation}
f_{Q}\left[I^{\text{tot}}\left(x,t\right)\right]=\frac{1}{I^{\text{tot}}\left(x,t\right)\sigma_{Q}\sqrt{2\pi}}\exp\left[-\frac{\left|\ln\left[I^{\text{tot}}\left(x,t\right)\right]-\mu_{Q}\right|^{2}}{2\sigma_{Q}^{2}}\right],\label{eq:fQ}
\end{equation}}
where $\mu_Q$ and $\sigma_Q$ are the mode and scale of the natural logarithm of $I^{\text{tot}}\left(x,t\right)$, respectively.

\subsection{Simulation Experiments}
Our aim is to explore the effects of STPP on the tracking dynamics of neuronal populations by using CANNs. To study the dynamics of a CANN with STPP, we built a model using the equations described in the previous subsection and conducted simulation experiments.
\subsubsection{Building the Model}
\begin{figure}[h!]
	\begin{center}
		\includegraphics[width=17.5cm]{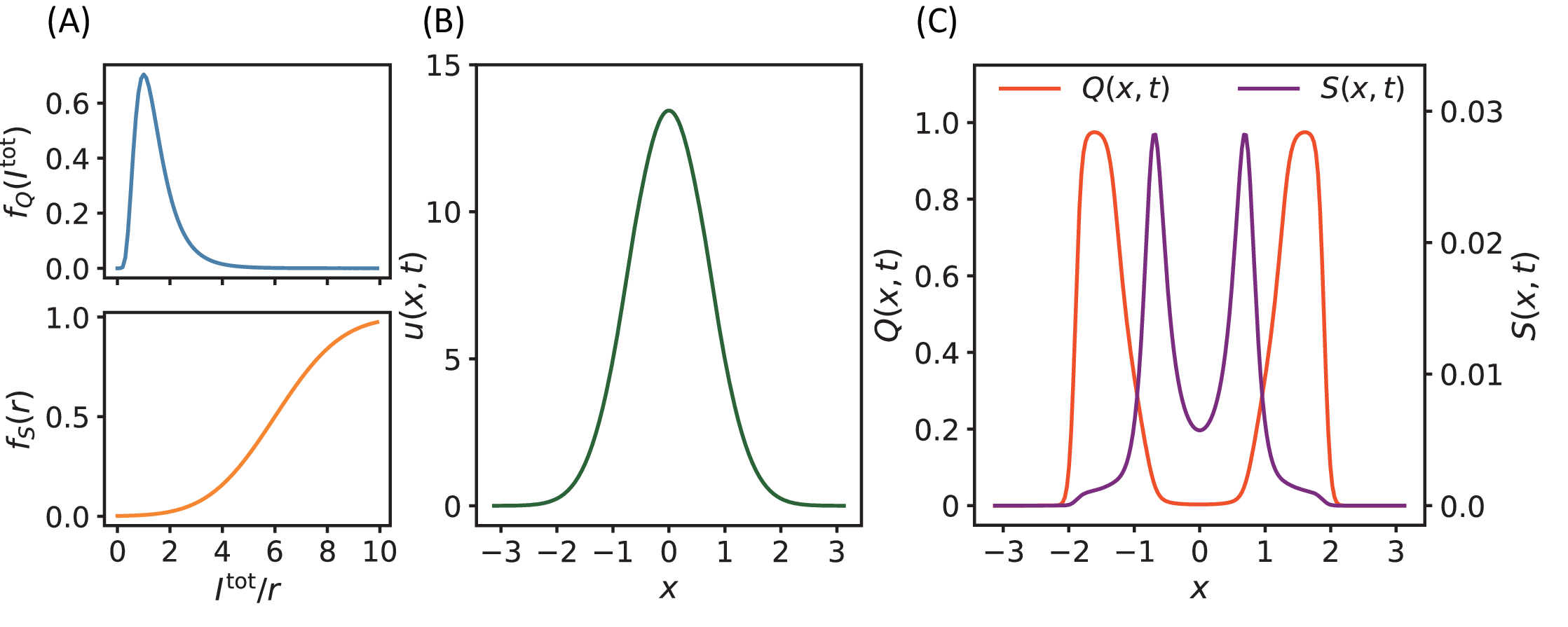}
	\end{center}
        \caption{(A) Profiles of $f_Q$, $f_S$, where $\sigma_{Q}=0.5$, $\mu_{Q}=0.25$ for $f_{Q}$, and $r_{0}=6$, $\sigma_{S}=2$ for $f_{S}$. (B) As shown, the {synaptic input} $u\left(x,t\right)$ of neurons is almost Gaussian, where the position of the external input is at 0. (C) The corresponding latent and enhancing modulations $Q\left(x,t\right)$ and $S\left(x,t\right)$ of $u\left(x,t\right)$ in (B). As shown, $Q\left(x,t\right)$ and $S\left(x,t\right)$ are initially symmetric with respect to the center of $u\left(x,t\right)$ when there is no external moving drive to $u\left(x,t\right)$. Parameters of (B) and (C): $a=0.5$, $k=0.5$, $A=3.0$, $\alpha=0.02$, $\beta=0.1$.}\label{fig:fig2}
\end{figure}

For $f_Q$ in Equation \eqref{eq:fQ} and $f_S$ in Equation \eqref{eq:fS}, we assigned the following empirical values: $\sigma_{Q}=0.5$, $\mu_{Q}=0.25$, $r_{0}=6$ and $\sigma_{S}=2$. As shown in Figure \ref{fig:fig2}A, $f_Q$ is log-normally distributed, with the latent modulation $Q$ increasing most strongly when the total input $I^{\text{tot}}\left(x,t\right)$ is relatively weak. $f_S$ is sigmoid-shaped, enabling $S\left(x,t\right)$ to be increased at strong membrane potential. Here, the membrane potential is represented by $r\left(x,t\right)$ \citep{Latimer2019}. The profile of the {synaptic input} $u\left(x,t\right)$ is shown in Figure \ref{fig:fig2}B, which is Gaussian shaped where the center of the external input $I^{\text{ext}}\left(x,t\right)$ (not plotted) is at 0. The corresponding $Q\left(x,t\right)$ and $S\left(x,t\right)$ are initially symmetric with respect to the center of $u\left(x,t\right)$ when there is no translational separation between $I^{\text{ext}}\left(x,t\right)$ and $u\left(x,t\right)$ (Figure \ref{fig:fig2}C).

In our simulation experiments, we took the head directions as the example stimuli. Hence, $x$ is restricted in the space of $-\pi < x\le\pi$, and the distance between $x$ and $x^\prime$ is in a periodic condition that is calculated by
\begin{equation}
    \text{dist}\left(x,x^\prime\right)=\begin{cases}
x-x^\prime+2\pi,&\text{if}\left(x-x^\prime\right) \le-\pi, \\
x-x^\prime,&\text{if}-\pi < \left(x-x^\prime\right)\le\pi,\\
x-x^\prime-2\pi,&\text{if}\left(x-x^\prime\right)>\pi.
\end{cases}\label{dist}
\end{equation}
{ One should also note that the model we used is a re-scaled model \citep{Fung2012NC}. Therefore, the number of neurons is not a factor determining the behavior of the system.}

\subsubsection{Tracking the External Stimulus}

To see how the network reacts to a stimulus, we set the external input to be
\begin{equation}
    I^{\text{ext}}\left(x\right)=A\exp\left(-\frac{\left|x-z_0\right|^2}{4a^2}\right),\label{Iext}
\end{equation}
where $A$ is the magnitude of the input. $z_0$ is the stimulus position. When considering a continuously moving stimulus, $I^{\text{ext}}\left(x,t\right)$ is defined as
\begin{equation}
    I^{\text{ext}}\left(x,t\right)=A\exp\left[-\frac{\left|x-z_0\left(t\right)\right|^2}{4a^2}\right].\label{Iext}
\end{equation}
Without loss of generality, we considered the stimulus position at time $t=0$ to be $z_0=0$, and the stimulus to move at a constant angular velocity $v_{\text{ext}}$ thereafter, i.e., $z_0\left(t\right)=v_{\text{ext}}t$.

With regard to the tracking behavior simulations, we altered the strength of STPP by adjusting $\alpha$ and $\beta$ and altered the external factors by adjusting $v_{\text{ext}}$ and $A$ to see how these factors influence the network's tracking performance. To measure the network's tracking performance when it is exposed to a continuously moving stimulus, we used the displacement between the network state and the stimulus position as an index. Note that the network state trails behind the stimulus when $s$ and $v_{\text{ext}}$ have opposite signs, whereas it predicts the position of the stimulus when they have the same sign. The displacement is given by $s=z\left(t\right)-z_0\left(t\right)$, where $z\left(t\right)$ is the center of mass of $u\left(x,t\right)$, which is calculated by
\begin{align}
z\left (t\right ) & = \tilde{x}+
\frac{\int dx \text{dist}\left(x,\tilde{x}\right)
 u\left(x,t\right)
}{\int dxu\left (x,t\right) },\\
\tilde{x} &=\underset{x}{\operatorname{arg\,max}}\;u\left (x,t\right).
\end{align}

\subsubsection{Measuring the Intrinsic Speed and the Anticipatory Time}

A study \citep{Fung2015} on dynamical behaviors in neural fields suggested that the models with inhibitory feedback modulations could support spontaneously moving profiles without any persistent external input. Therefore, we asked the following question: in the absence of external input, does a CANN with STPP have intrinsic motion? If the answer is yes, the follow-up question arises: is intrinsic motion caused by STPP? Note that without external input, when the network state becomes translationally unstable, it moves with a constant speed, i.e., intrinsic speed \citep{Fung2015}. To investigate the intrinsic dynamics of a CANN with STPP, we measured the intrinsic speeds, denoted as $v_{\text{int}}$, of the models when the STPP strength was varied. In the simulations, we first let the system reach its stationary state. After that, we removed the external input and manually shifted $u$ in the direction of positive $x$ by $2\pi / 200$ for every $\tau_\text{s}$. After 100 $\tau_\text{s}$, we terminated all manual intervention and let the system evolve. When the motion of the system state became steady, we recorded the speed of the intrinsic motion as $v_{\text{int}}$. For the STPP regimes, both $\alpha$ and $\beta$ were selected from 0 to 0.2 with a step of 0.004.

By analyzing the dynamical properties of the system, the study \citep{Fung2015} also found that intrinsic motion is the internal drive of anticipation. Hence, after obtaining the intrinsic speeds of the CANNs under various STPP regimes, we considered the following question: if intrinsic motion is present, is it also the internal drive of anticipatory behavior in our model? Therefore, to investigate the causality of anticipation, we measured the maximal anticipatory time $T_\text{ant}$ of the CANNs under the same STPP regimes as those used in the simulations for intrinsic speeds. Here, the anticipatory time $\tau_\text{ant}$ is defined as $s/v_\text{ext}$, which is a constant when the moving bump is steady. $T_\text{ant}$ is the maximum of $\tau_\text{ant}$ over a range of $v_\text{ext}$ under a given STPP regime. In the simulations, $\alpha$ and $\beta$ were also selected from 0 to 0.2 with a step of 0.004. $v_\text{ext}$ was altered from 0 to 0.008 rad/ms with a step of 0.0002. The magnitude $A$ of the stimulus is 3.0.

\section{Results}
\subsection{Tracking Behaviors}
To observe the tracking behaviors of the CANNs with STPP, we visualized the outlines of the network's motions, first in the presence of an abruptly changing stimulus and then in the presence of a continuously moving stimulus.
\subsubsection{In the Presence of an Abruptly Changing Stimulus}
\begin{figure}[h!]
        \begin{center}
            \includegraphics[width=14cm]{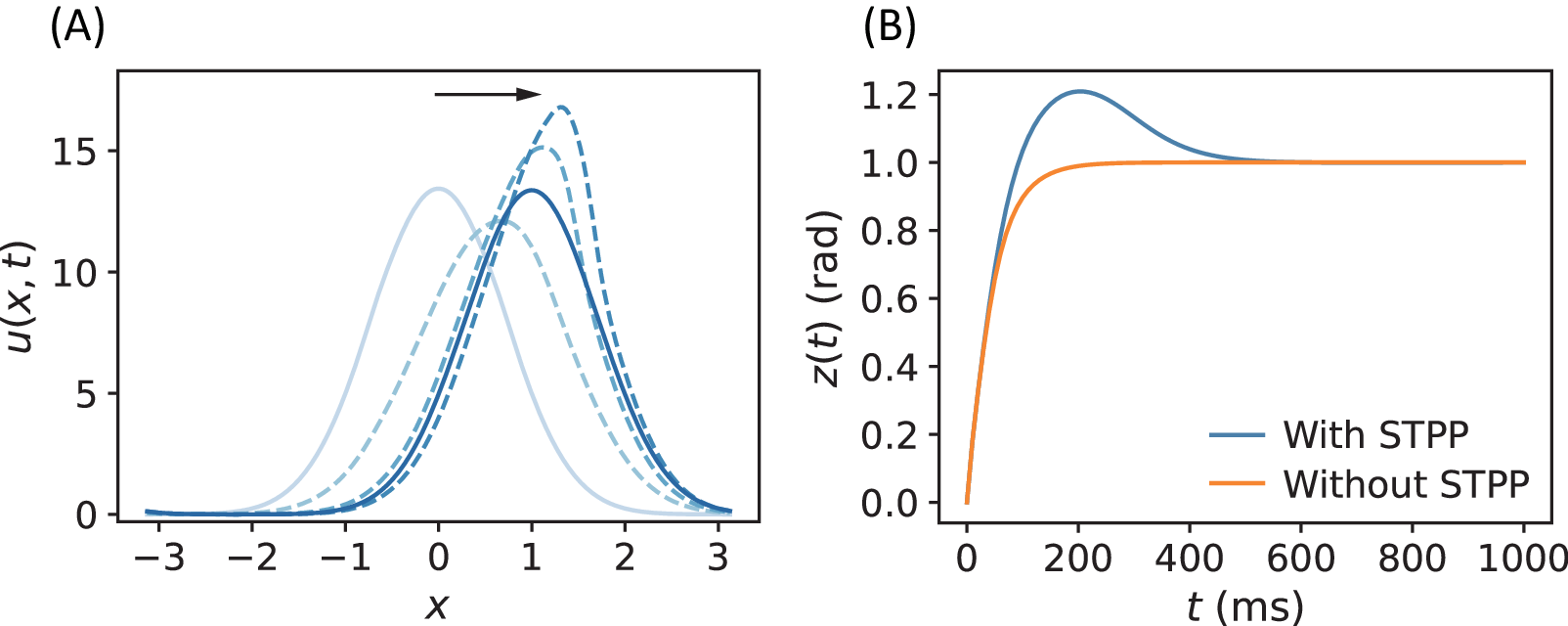}
        \end{center}
        \caption{(A) Snapshot of $u\left(x,t\right)$ of a CANN with STPP when the stimulus abruptly changes from $z_0=0$ to $z_0=1$. The direction of the motion of $u\left(x,t\right)$ follows the black arrow, and the blue color gradually darkens, representing the new phases over time. The corresponding $z\left(t\right)$ of $u\left(x,t\right)$ is shown in (B). Tracking of an abruptly changing stimulus shows prediction in the CANN with STPP compared to that without STPP. Parameters: $a=0.5$, $k=0.5$, $A=3.0$, model with STPP: $\alpha=0.02$, $\beta=0.1$, model without STPP: $\alpha=0$, $\beta=0$.}\label{fig:fig3}
\end{figure}

We first compared the tracking pattern of a CANN with STPP ($\alpha=0.02$, $\beta=0.1$) to that without STPP ($\alpha=0$, $\beta=0$) when the stimulus abruptly changed its position $z_0$. The stimulus position was initially at $z_0=0$ and then shifted to $z_0=1$. As expected, the {synaptic input} $u\left(x,t\right)$ moved to align the center of mass $z$ with the stimulus position $z_0$ (Figure \ref{fig:fig3}A). Further, the snapshot shows an overshoot of the destination, which is a sign of translational instability of the CANN with STPP. A clearer comparison of $z$ was made between a network with STPP and that without STPP (Figure \ref{fig:fig3}B). This instability indicates the potential of STPP for anticipatory tracking. 
\subsubsection{In the Presence of a Continuously Moving Stimulus}
\begin{figure}[h!]
        \begin{center}
            \includegraphics[width=16cm]{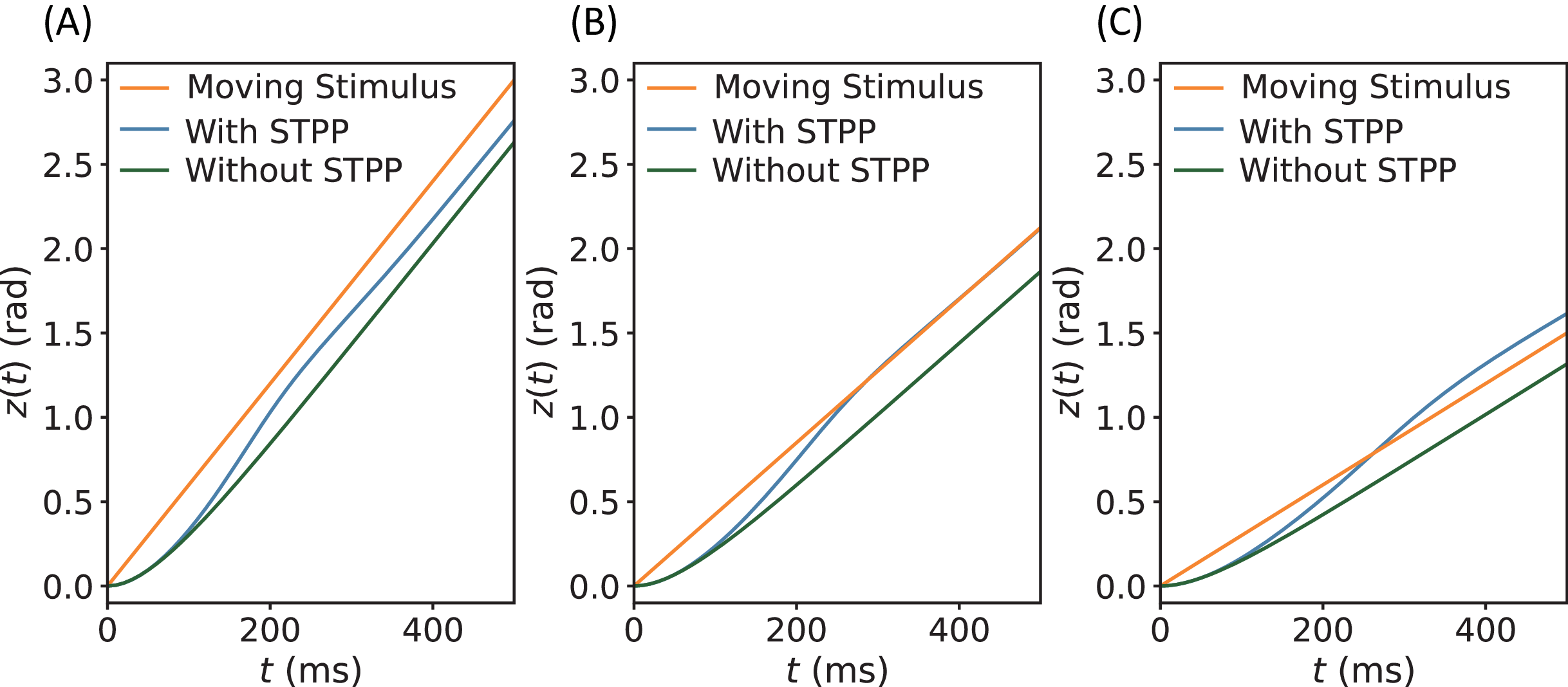}
        \end{center}
        \caption{Three comparisons of trackings of a continuously moving stimulus in CANNs with and without STPP. Networks without STPP always exhibit a delay to the moving stimulus, whereas networks with STPP lead with respect to the stimulus. For the same network with STPP, there are mainly three cases of the tracking depending on the angular velocity of stimulus $v_{\text{ext}}$: (A) delayed tracking when $v_{\text{ext}}=0.006$ rad/ms; (B) perfect tracking with zero lag when $v_{\text{ext}}=0.00425$ rad/ms; and (C) anticipatory tracking with constant time when $v_{\text{ext}}=0.003$ rad/ms. Parameters: $a=0.5$, $k=0.5$, $A=2.0$, models with STPP: $\alpha=0.02$, $\beta=0.1$, models without STPP: $\alpha=0$, $\beta=0$.}\label{fig:fig4}
\end{figure}

Next, we explored the tracking behaviors of the networks in the presence of a continuously moving stimulus by allowing the stimulus to shift with a constant angular velocity $v_\text{ext}$. The CANNs were able to track the moving stimuli by using an approximate speed, which was consistent with how the HD cell system tracks rotating visual cues \citep{Ajabi2023}. The simulations show that the networks without STPP always trailed behind the stimulus regardless of the speed. However, after STPP was applied, three cases of tracking patterns were mainly observed, depending on $v_\text{ext}$: (1) delayed tracking (Figure \ref{fig:fig4}A), (2) perfect tracking with zero lag (Figure \ref{fig:fig4}B), and (3) anticipatory tracking with a constant lead (Figure \ref{fig:fig4}C). Notably, even when there was a delay, the CANN with STPP performed better in tracking the stimulus than pure CANN. In our cases, delayed, perfect and anticipatory trackings occurred in order from fast to slow $v_\text{ext}$. However, delay and alignment may occur in both slow and fast zones, which is discussed in the next subsection. These results suggest that the predictions of the CANNs with STPP depend on the speeds of the stimulus.
\subsection{Anticipatory Tracking}
\begin{figure}[h!]
	\begin{center}
		\includegraphics[width=17.5cm]{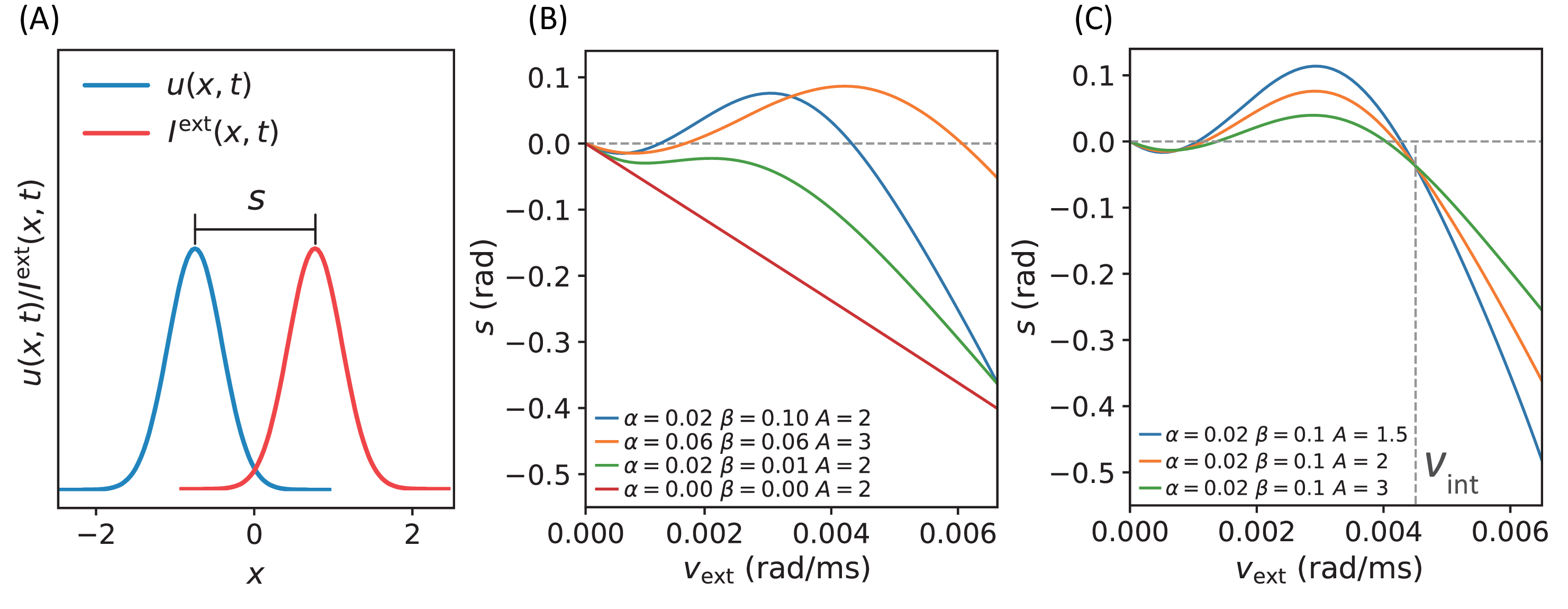}
	\end{center}
	\caption{(A) The displacement $s$ between the network state and stimulus position is defined by $s=z\left(t\right)-z_0\left(t\right)$, which indicates anticipation when $s$ and $v_{\text{ext}}$ have the same signs and delay when $s$ and $v_{\text{ext}}$ have opposite signs. (B) Comparisons of $s$ of the CANNs at varying speeds $v_{\text{ext}}$ of moving stimulus with different STPP parameters ($\alpha$, $\beta$). As shown, the network without STPP ($\alpha=0.00$, $\beta=0.00$) linearly lags behind the stimulus (grey dashed line at $s=0$) as $v_{\text{ext}}$ increases. The network with weak STPP ($\alpha=0.02$, $\beta=0.01$) also lags behind the stimulus. However, the networks with certain strengths of STPP, e.g., $\alpha=0.06$, $\beta=0.06$ and $\alpha=0.02$, $\beta=0.10$, predict the motion of the stimulus within a limited range of $v_{\text{ext}}$. With respect to the same network ($\alpha=0.02$, $\beta=0.10$), the dependency of the anticipatory performance on the intensity of the stimulus ($A$) is shown in (C). These networks have similar patterns, but the weaker the intensity, the greater the anticipation and the larger the velocity range. Moreover, it is notable that these curves converge on a point at a specific velocity where the displacement is independent of the intensity of the stimulus. The velocity of the confluence point is called intrinsic speed $v_{\text{int}}$ of the network. Parameters: $a=0.5$, $k=0.5$.}\label{fig:fig5}
\end{figure}

Next, we examined the dependencies of anticipatory performance on internal (i.e., STPP parameters) and external (i.e., the strength of the stimulus) factors. The displacement $s$ is a measure of the tracking performance (see the section titled "Simulation Experiments" for the equation), as shown in Figure \ref{fig:fig5}A. First, we investigated how the STPP parameters affect the model's tracking performance. We compared $s$ among different STPP regimes, the results of which indicate that anticipatory tracking is achieved in a certain range of STPP strength. As shown in Figure \ref{fig:fig5}B, the network without STPP ($\alpha=0$, $\beta=0$) linearly lagged behind the stimulus (the grey dashed line at $s=0$). Moreover, when the STPP was relatively weak due to the small scale of parameters, e.g., the dark green line ($\alpha=0.02$, $\beta=0.01$), the network failed to make predictions although it showed less lag. In contrast, the networks with strong STPP elicited prediction successfully over a considerable range of angular velocities. For instance, the network with $\alpha=0.02$ and $\beta=0.1$ could anticipate stimuli with velocities ranging from 69$^\circ/\text{s}$ to 240$^\circ/\text{s}$. A stronger STPP regime ($\alpha=0.06$, $\beta=0.06$) facilitated anticipation of a broader range of velocities ranging from 92$^\circ/\text{s}$ to 338$^\circ/\text{s}$. Interestingly, we found that at slower velocities, the networks performed with a tiny delay, and the threshold of the velocity for anticipation was related to both the STPP regime and the stimulus (Figure \ref{fig:fig5}B, C). These results reveal the possible implications of STPP for prediction. On the one hand, effective anticipation of stimuli over an extensive range of velocities gives the neural system great flexibility to adapt to a varying stimulus. On the other hand, the correspondence between the range of anticipation-achievable velocities and the diverse intensities of STPP may imply that distinct brain regions or neurons equipped with distinct synaptic plasticity enable diverse anticipatory tracking performance levels.

Second, to understand how the strength of a stimulus affects anticipatory performance, we applied different magnitudes $A$ of a stimulus to the same network and compared the results. According to the results shown in Figure \ref{fig:fig5}C, the anticipatory lead was greater and the anticipatory velocity range was larger under the weaker stimulus. Moreover, these traces were confluent at a specific velocity, where the displacement was independent of the intensity of the stimulus. Interestingly, this velocity was the same as the intrinsic speed $v_\text{int}$ of the network, which is described and discussed in the next subsection. In line with the results of natural tracking obtained in an earlier study \citep{Fung2015}, our results also showed confluent behavior, which reveals a certain relationship between intrinsic dynamics and tracking dynamics. That is when the stimulus speed is the same as the intrinsic speed, the displacement is related only to the network itself.
\subsection{Intrinsic Dynamics and Anticipatory Time}
\begin{figure}[h!]
	\begin{center}
		\includegraphics[width=17.5cm]{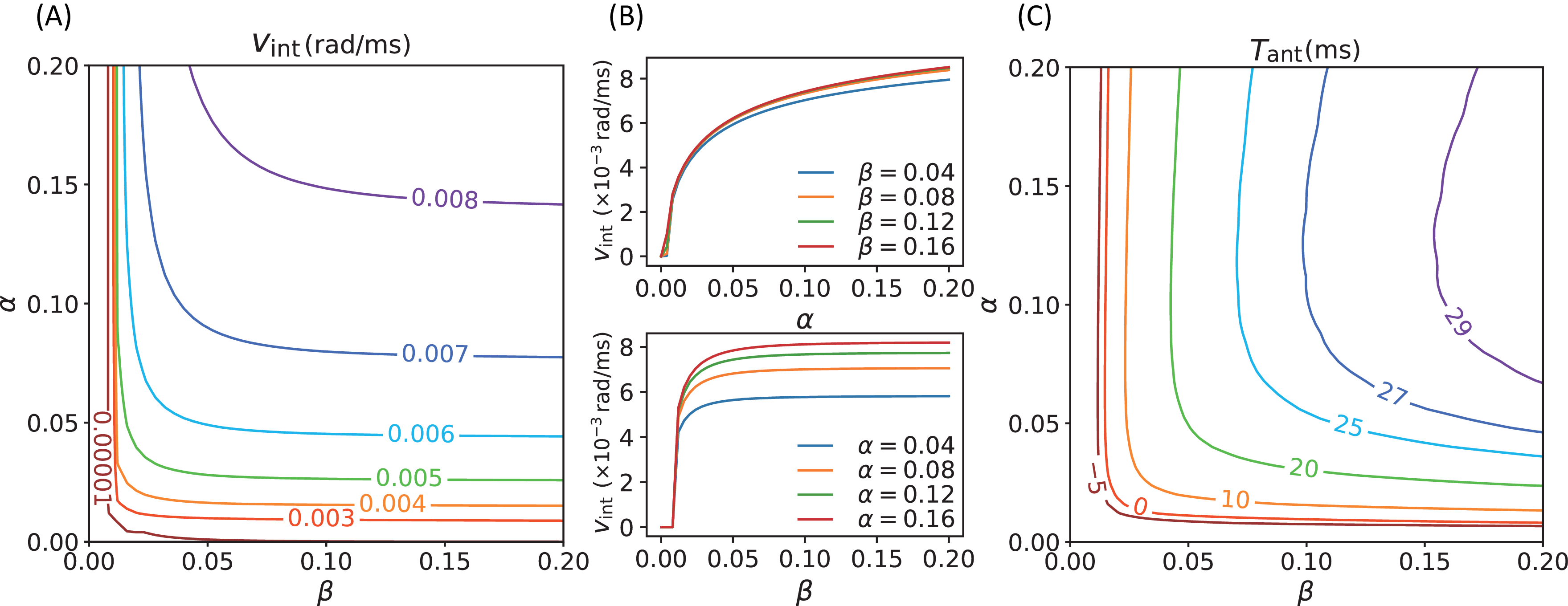}
	\end{center}
        \caption{(A) Contour map of intrinsic speeds $v_{\text{int}}$ of CANNs with different STPP regimes. $v_{\text{int}}$ increases as $\alpha$ or $\beta$ (or both) increases (increase). (B) However, $v_{\text{int}}$ increases notably along $\alpha$ when $\beta$ is fixed (top), whereas it becomes almost constant after $\beta$ reaches a certain level when $\alpha$ is fixed (bottom). Therefore, $v_{\text{int}}$ is not clearly distinguishable with respect to $\beta$ whereas it exhibits a clearly hierarchical response to $\alpha$. It seems that $v_{\text{int}}$ is more sensitive to $\alpha$ than to $\beta$. (C) Contour map of the maximal anticipatory time $T_{\text{ant}}$ of CANNs with the same STPP regimes as those of (A). All of the stimuli possess the same magnitude $A=3.0$, and  $v_{\text{ext}}\in\left[0,0.008\right]$ rad/ms. $T_{\text{ant}}<0$ means delays for all $v_{\text{ext}}$. The anticipation appears after $v_{\text{int}}>0$ and $T_{\text{ant}}$ exhibits a similar trend as $v_{\text{int}}$, which suggests that intrinsic motion is an internal drive of anticipation and is a necessary condition for it probably. Parameters: $a=0.5$, $k=0.5$.}\label{fig:fig6}
\end{figure}

To work out the relationship between the intrinsic dynamics and the tracking dynamics of the network, we next explored the intrinsic dynamics in the absence of a persistent stimulus (we had already tested the tracking dynamics in its presence). The contour map of the intrinsic speeds $v_\text{int}$ under various combinations of $\alpha$ and $\beta$ is showed in Figure \ref{fig:fig6}A. $v_\text{int}$ increased as $\alpha$ or $\beta$ (or both) increased. Here, $v_\text{int}\le0.00001$ can be considered static because 0.00001 rad/ms is approximately equal to 0.57 $^\circ/\text{s}$, which is much lower than the majority of the velocities we recorded. In the region where $v_\text{int}\le0.00001$, the network became practically static after manual intervention was terminated. In the moving region, the network became translationally unstable and moved with an intrinsic speed. The intrinsic motion was induced by the strong STPP, whereas the static region was located in parameter regions that lacked STPP or were under weak STPP regimes. Figure \ref{fig:fig6}B shows the separate effects of $\alpha$ and $\beta$. $v_\text{int}$ increased notably along $\alpha$ when $\beta$ was fixed, whereas when $\alpha$ was fixed, it became almost invariant after $\beta$ reached a relatively high value. Combining the top and bottom sub-figures of Figure \ref{fig:fig6}B reveals that $v_{\text{int}}$ was not clearly distinguishable with respect to $\beta$ but exhibited a clearly hierarchical response to $\alpha$. This behavior implies that $v_\text{int}$ is more sensitive to $\alpha$ than to $\beta$.

A certain relationship between the intrinsic dynamics and the tracking dynamics was shown in the previous subsection. In view of this, the question arises: is intrinsic motion the internal drive of anticipatory behavior? To answer this question, we measured the maximal anticipatory time $T_\text{ant}$ of the CANNs under the same STPP regimes as those in Figure \ref{fig:fig6}A to understand the causality of the anticipation. As shown in Figure \ref{fig:fig6}C, the anticipation ($T_\text{ant}>0$) appeared after $v_\text{int}>0.00001$. Also, the contour map of $T_\text{ant}$ and that of the intrinsic speeds shared a similar trend. These results indicate that stimulus prediction occurs only when there is intrinsic motion and the network is not in an internally static state. Hence, we assume that STPP-induced intrinsic motion is an internal drive of anticipation and a necessary condition for it. The pattern of the maximal anticipatory time map does not perfectly match the intrinsic speed map owing to the nontrivial influence of external input.
\subsection{Analysis of Translational Stability of the System}
\begin{figure}[h!]
	\begin{center}
		\includegraphics[width=17.5cm]{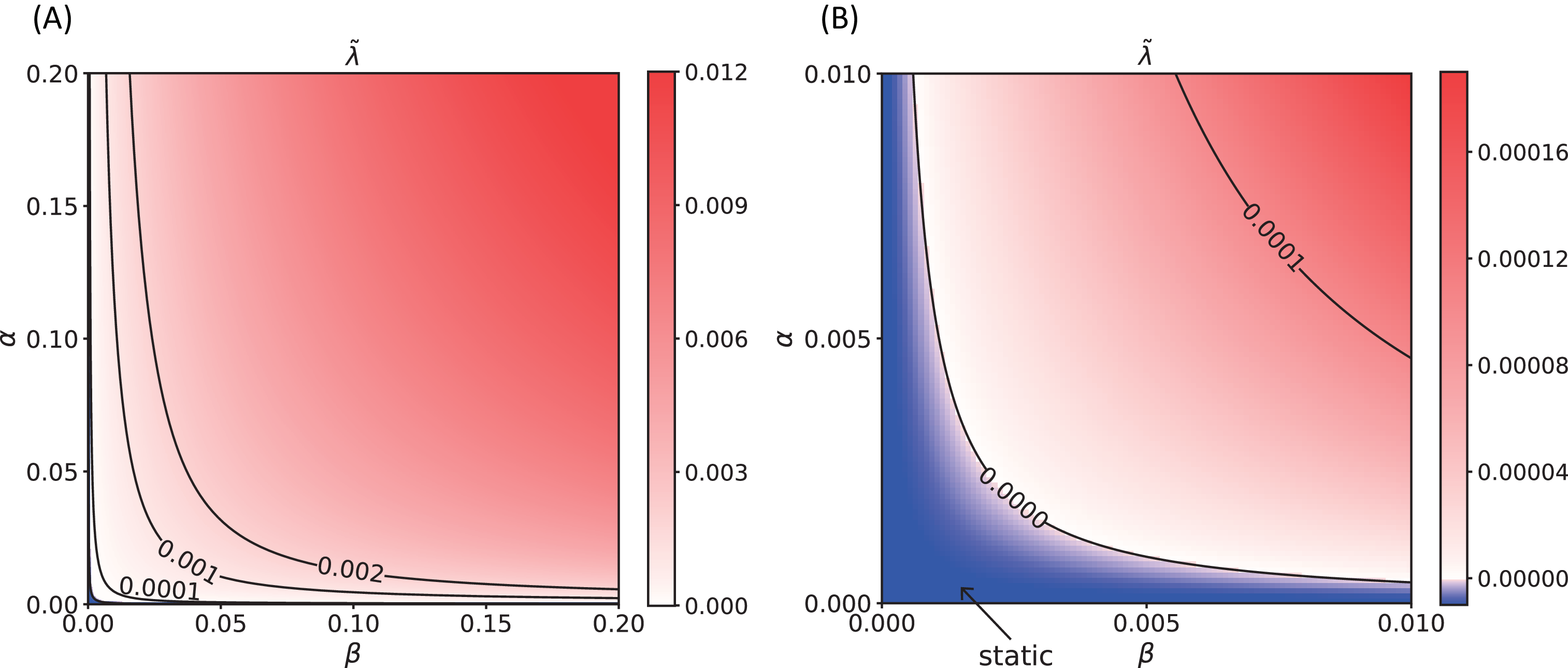}
	\end{center}
	\caption{(A) Heatmap of the translational stability analysis results, which shares a similar pattern to that of the intrinsic speeds. The unit of $\alpha$ and $\beta$ is 0.001. (B) Enlargement of the bottom left corner of (A). The unit of $\alpha$ and $\beta$ is 0.0001. The blue region is the static phase of the network where the maximal eigenvalue $\tilde\lambda\le0$. Elsewhere, it is unstable and spontaneously moves when there is a tiny perturbation. In particular, the parameter region of the intrinsic motion is located in the moving phase of the system. This analysis suggests that the intrinsic dynamics of the network are caused by the translational instability of this system.}\label{fig:fig7}
\end{figure}

In a dynamical system, spontaneous movement without an external input occurs when the static solution becomes unstable to positional displacement in some parameter regions. To understand the intrinsic dynamics of a CANN with STPP, we studied the translational stability of static solutions of our model. We considered network states with a positional displacement to be
\begin{align}
u\left(x,t\right) & =u_{0}\left(x\right)+u_{1}\left(t\right)\frac{du_{0}\left(x\right)}{dx},\\
S\left(x,t\right) & =S_{0}\left(x\right)+S_{1}\left(t\right)\left(x-z\right)S_{0}\left(x\right),\\
Q\left(x,t\right) & =Q_{0}\left(x\right)+Q_{1}\left(t\right)\left(x-z\right)Q_{0}\left(x\right),
\end{align}

where $u_{0}\left(x\right)$, $S_{0}\left(x\right)$, and $Q_{0}\left(x\right)$ and $u_{1}\left(t\right)$, $S_{1}\left(t\right)$, and $Q_{1}\left(t\right)$ are the stationary states and the displacements of $u\left(x,t\right)$, $S\left(x,t\right)$, and $Q\left(x,t\right)$, respectively. $z$ is the center of mass of $u_{0}\left(x\right)$. As derived in Supplementary Material,
\begin{equation}
\frac{d}{dt}\left(\begin{array}{c}
u_{1}\left(t\right)\\
S_{1}\left(t\right)\\
Q_{1}\left(t\right)
\end{array}\right)=\left(\begin{array}{ccc}
M_{uu} & M_{uS} & 0\\
M_{Su} & M_{SS} & M_{SQ}\\
M_{Qu} & 0 & M_{QQ}
\end{array}\right)\left(\begin{array}{c}
u_{1}\left(t\right)\\
S_{1}\left(t\right)\\
Q_{1}\left(t\right)
\end{array}\right).
\label{eq:solution}\end{equation}
For the dynamical system described by Equation \eqref{eq:solution}, the eigenvalue $\lambda$ of the matrix composed of $M_{\psi\varphi}$ ($\psi,\varphi\in \left\{u, S, Q\right\}$) and 0 determines its stability. We calculated the eigenvalues of networks with varying STPP regimes, letting the maximum of the eigenvalues for each regime be denoted as $\tilde\lambda$. In the static phase, $\tilde\lambda\le0$, the network states are stable and static. In the moving phase, $\tilde\lambda>0$, the system would diverge due to positional displacement, i.e., perturbation, leading to spontaneously moving bumps. As illustrated in Figure \ref{fig:fig7}, the network states were static when $\tilde\lambda\le0$ in some STPP regimes, specifically weak STPP regimes. Otherwise, the network state bump moved spontaneously when there was interference. That is, STPP increases the translational instability, i.e., the mobility, of the CANNs. Remarkably, the parameter region of intrinsic motion is in the moving phase of the system. The heatmap of $\tilde\lambda$ has a similar pattern to that of the contour map of the intrinsic speeds shown in Figure \ref{fig:fig6}A. Overall, the results indicate that the intrinsic motion of the network is caused by the translational instability of this system, which is induced by STPP.

\section{Discussion}
{ In this study, we implemented the feedforward signal amplification mechanism of STPP in CANNs. In the model, the STPP effect is modeled by adding two dynamic functions ($Q\left(x,t\right)$ and $S\left(x,t\right)$) to the original CANN \citep{Wu2005}.} {$Q\left(x,t\right)$ abstractly models the proportion of NDMARs that are in the bound-but-blocked state, while $S\left(x,t\right)$ abstractly models the temporal enhancement dependent on the opening of NMDARs from the bound-but-blocked state.}
{To trigger the bound-but-blocked state, a sufficient but limited synaptic input should be applied (Figure 1B in \citep{Yang2016}). To model this effect, we chose the function evoking $Q\left(x,t\right)$, i.e., $f_Q$, to be a bump-shaped log-normal distribution function. On the other hand, when the input to the postsynaptic neuron is uncommonly large, the postsynaptic membrane potential can be sufficient to remove magnesium ions \citep{Jahr1990, VargasCaballero2004}. As a result, the evoked postsynaptic current becomes stronger than that with NMDARs blocked \citep{Yang2016}. To model this magnesium removal efficacy, we chose the cumulative distribution function $f_S$ to trigger the transition from the bound-but-blocked state to the open state. As the average membrane potential can be represented by a firing rate when reasonably assuming that the below-threshold membrane potentials of the postsynaptic neurons are normally distributed around the average value \citep{Latimer2019}, the cumulative distribution function of the firing rate ($r\left(x,t\right)$) is an appropriate choice to model magnesium removal efficacy on a population of neurons sharing the same preferred stimulus. In summary, functions $f_Q$ and $f_S$ motivated by our previous evidence help visualize the STPP effect. 

The additional term $\left[1+S\left(x,t\right)\right]$ in Equation \eqref{eq:dudt} initiates concerns about its appropriateness due to different time constants of AMPARs and NMDARs. Even though the decay time constants of AMPAR and NMDAR are different, the STPP modulation takes into effect when NMDARs are primed and glutamates are re-released. When the opening of NMDARs and AMPARs are simultaneous, the whole evoked postsynaptic currents can be modulated by STPP.}
{ Moreover, $\left[1+S\left(x,t\right)\right]$ models the efficacy of presynaptic neuronal activity in the evocations of postsynaptic input. If a neuron was primed, the following information-bearing glutamate release could induce a surplus depolarization due to NMDAR-dependent STPP. Therefore, the modulation is on the evocation of postsynaptic response, not the synaptic current corresponding to receptors of particular types. However, the inhibition in this model is divisive. Nonetheless, the modulation still acts on the cortical and hippocampal networks in the preserved excitatory and inhibitory circuits, as in our previous study \citep{Yang2016}. Although we model the modulating effect on the excitatory synaptic currents, this modeling setting is sufficient to mimic the effect of STPP on CANNs.}

{Our simulation experiments showed that the implementation of STPP in CANNs successfully enabled anticipatory tracking of a moving stimulus and we also explored the relationship of the model dynamics and anticipation.} First, we tested the tracking behaviors of CANNs with STPP when they were exposed to an abruptly changing stimulus and a continuously moving stimulus, and found that this model could predict the future locations of the stimulus. Second, we investigated the intrinsic dynamics of the networks with varying STPP regimes; the results showed that a certain level of STPP supported the intrinsic motion in this model {and the intrinsic speed was more sensitive to $\alpha$.} Interestingly, when the angular velocity of the moving stimulus reached the intrinsic speed of the given model, the tracking delay was independent of the strength of the stimulus. Third, by comparing the pattern of the maximal anticipatory time with that of the intrinsic speeds, we found that intrinsic motion is the internal drive of anticipation. Lastly, we analyzed the translational stability of the stationary states of our networks with different STPP parameters. The results implied that strong STPP enhanced the mobility of the system and thus induced intrinsic motion. {Taking all the results into account, we noticed that strong $\alpha$ (opening rate of NMDARs) and $\beta$ }{(transition rate of NMDARs from the glutamate-unbound state to the bound-but-blocked state) enable anticipation with broader coverage of stimulus velocities and larger maximal anticipatory time. Physiologically, the transition rate of NMDARs to the bound-but-blocked state reflects how fast and efficiently glutamates bind to NMDARs, which is basically dependent on glutamate-NMDAR binding rate and affinity \citep{Paoletti2007, Singh2011}. Generally, these factors are synaptic specific. Intuitively, the fast opening of NMDARs can help neurons react expeditiously for stimulus chasing, and fast binding can prime NMDARs promptly for further opening. Given that the opening rate (or rise time), the binding rate and affinity are associated with the subunits of NMDAR \citep{Paoletti2007, Singh2011, Tu2014}, natural differences in NMDAR subtypes \citep{Buller1994, Paoletti2013} would possibly lead to diverse tracking performance for stimuli in various brain regions. In contrast, an abnormal subunit expression of NMDARs may impair perception by the failure of time delay compensation. Collectively, our findings indicate that NMDAR-dependent STPP may underlie sensory prediction, effectively compensating for neural delay and efficiently supporting sensorimotor control.}

\begin{figure}[h!]
	\begin{center}
		\includegraphics[width=17.5cm]{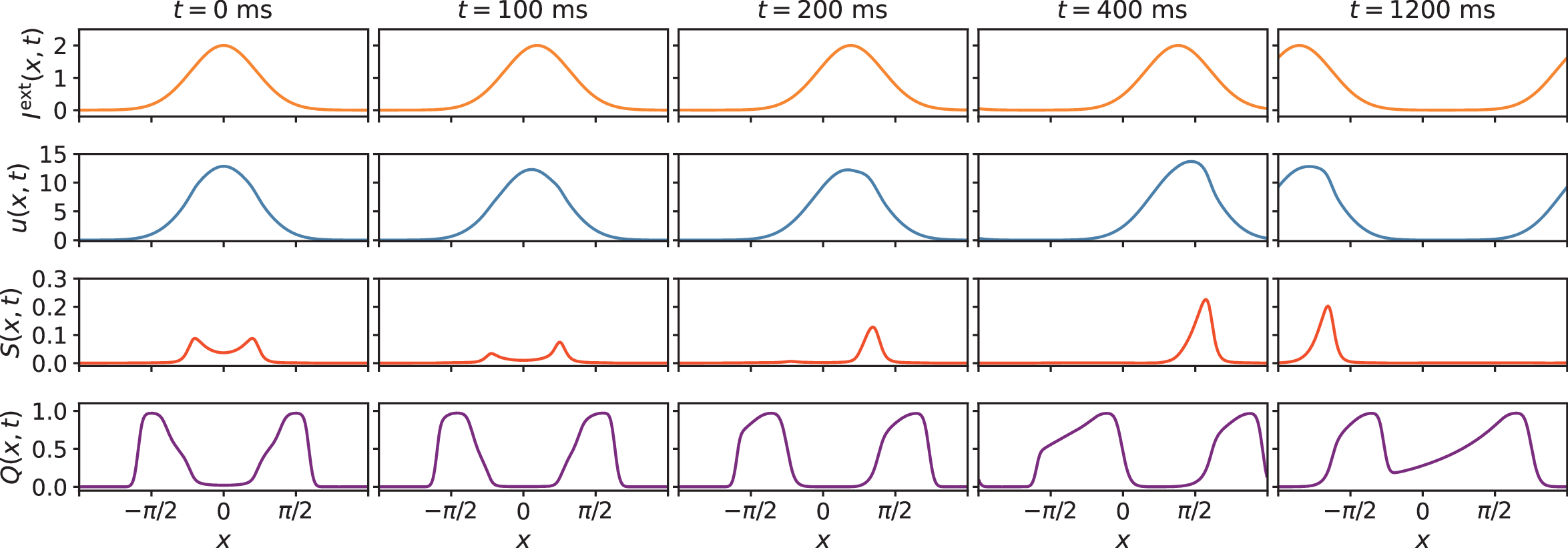}
	\end{center}
        \caption{Extracted eight states of $I^{\text{ext}}\left(x,t\right)$, $u\left(x,t\right)$, $S\left(x,t\right)$, and $Q\left(x,t\right)$ of the network. At the beginning of the external input movement, $u\left(x,t\right)$, $S\left(x,t\right)$ and $Q\left(x,t\right)$ undergo distortions. The biased distortion of $Q\left(x,t\right)$ induces a unimodal distribution of the enhancing modulation ($S\left(x,t\right)$) with a peak on neighboring neurons that prefer future stimulus positions, thus skewing $u\left(x,t\right)$ toward future positions. This asymmetry results in stronger activation of neurons that favor future positions, which enables this neural system to sense future events. Parameters: $a=0.5$, $k=0.5$, $A=2.0$, $\alpha=0.02$, $\beta=0.1$, $v_{\text{ext}}=0.003$ rad/ms.}\label{fig:fig8}
\end{figure}

The underlying mechanism of how STPP can facilitate anticipation can be intuitively understandable. In Figure \ref{fig:fig8}, eight frames of network states were extracted for visualization. From the time-varying states and Equations \eqref{eq:dudt}, \eqref{eq:r}, \eqref{eq:dSdt} and \eqref{eq:dQdt}, we can understand how anticipation occurs. When the stimulus starts moving, the biased distortion of $Q\left(x,t\right)$ induces a unimodal distribution of the enhancing modulation $S\left(x,t\right)$ with the peak on neighboring neurons that prefer future stimulus positions. This consequently skews $u\left(x,t\right)$ toward future positions. This asymmetry leads to stronger activation of the neighboring neurons that exceed the peak of the stimulus, thereby facilitating prediction. In this case, after around 1200 ms, the states became translationally stable with the same speed as the stimulus; only the locations (not shapes) of the states changed, and thus a constant anticipation time was maintained. Overall, the essence of the anticipatory phenomenon is rooted in the effect of NMDAR-dependent STPP. The information on stimulus positions can be latent in neighboring neurons that are less activated than the stimulus peak-aligned neurons, due to the small input. As the input shifts, the stronger input induces enhanced activation of the neighboring neurons that have stored information. This feedforward effect eventually skews the activation of neurons toward future positions and enables the neural system to sense future events.

Interestingly, despite being a response facilitation mechanism, STPP has the opposite function to that of short-term facilitation (STF). STF refers to a type of short-term synaptic plasticity that boosts the neuronal postsynaptic response as a result of the increased likelihood of neurotransmitter release due to the influx of calcium into the axon terminal after spike generation \citep{Zucker2002}. In a computational study \citep{Fung2012NC} of STF in CANNs, it was discovered that STF improved the behavioral stability of CANNs and served as a noise filter. STF can maintain strong activation of neurons in the active region in accordance with the stimulus position and strengthen interactions among neurons that are tuned to the stimulus. Unlike STPP, this stimulus-specific facilitation stabilizes the network, which makes it incapable of spontaneously moving to anticipate an external input. Inhibitory feedback modulations were considered the basic mechanisms of anticipatory tracking. The modulations include STD, which depresses the activation of postsynaptic neurons by depleting the available resources of neurotransmitters released from presynaptic neurons \citep{Fung2012}, and SFA, which refers to the reduction of neuronal excitability after prolonged stimulation \citep{Mi2014}. In both STD and SFA, the most active neurons receive the strongest negative feedback to counterbalance their responses, thereby increasing the probability that the locally active network state shifts to the continuously moving stimulus in a sequence. In contrast to inhibitory feedback modulations, which emphasize self-suppression by a feedback system, STPP spotlights the response facilitation on adjacent neurons in a feedforward manner to increase the mobility of the network and achieve anticipatory tracking. Moreover, STPP operates by utilizing the natural properties of NMDARs, which play a significant role in synaptic transmission, rather than relying on recurrent feedback systems, which may involve more auxiliary pathways and systems and may depend on repetitive or prolonged firing \citep{Zucker2002, Gutkin2014}. It appears that the STPP-dominant feedforward model is simpler and more energy-efficient for anticipatory tracking. 

Animal experimental results have suggested that the HD cell system can anchor to local cues, stably tracking rotating cues and maintaining traces when the cues are removed \citep{Taube1990, Goodridge1998, Zugaro2003, Ajabi2023}. However, the mechanism underlying the anchoring and trace maintenance remains unresolved. In 2023, Ajabi's team \citep{Ajabi2023} found a second dimension in addition to a single angular dimension, namely network gain, to help represent the HD cell system. This discovery indicates that additional information is necessary to fully comprehend how the system adapts to changeable sensory input although the origins of network gain are yet to be identified. They also reported that the system followed the cue's rotation speed to track it and that the system exhibited anticipation. Additionally, the system sustained the drift for some time after the cue was turned off. These findings support the idea that the HD cell system possesses an effective predictive tracking capacity and exhibits spontaneous drift based on experience. To some degree, our model achieved results that were similar to those obtained by these experiments. Our CANN-based model also tracked a moving stimulus with the same velocity \citep{Fung2010, Fung2015}. Moreover, the incorporation of STPP enabled our model to move spontaneously without persistent external input, i.e., it exhibited intrinsic motion, and showed the anticipatory tracking of a moving stimulus. A very early study \citep{Blair1997} reported a skew of the peak of the tuning curve toward future positions in the HD cell system rather than just a translational shift to future positions with an invariant bump shape, a finding that was interpreted as suggesting anticipation. This asymmetry can be induced by STPP in our model. As for the anticipatory time, different studies obtained varying average values such as 17 ms \citep{Blair1997}, 23 ms \citep{Taube1998}, and 25 ms \citep{Goodridge2000} in the anterior thalamus. The anticipatory time of HD cells can vary in different brain regions and even in different cells in the same region \citep{Blair1997, Taube1998, Goodridge2000}. In a large STPP parameter region, our model produced anticipations with maximal anticipatory time ranging from 0 to 30 ms, which were of the same order as that observed in animal experiments. Although STD \citep{Fung2012} and SFA \citep{Mi2014} can achieve tracking performance comparable to that obtained by us, the effect of NMDAR-based STPP should not be neglected.

In addition, to prove the robustness of our model to different values of the STPP time constant $\tau_2$, which represents the lifespan of latent information, we obtained contour maps of $v_{\text{int}}$ when $\tau_2=300$ ms, $\tau_2=400$ ms, $\tau_2=600$ ms, and $\tau_2=700$ ms through simulations. The results, which are illustrated in Supplementary Material Figure S1, show that the patterns and achievable ranges of intrinsic speeds in other conditions resemble those obtained with $\tau_2=500$ ms (Figure \ref{fig:fig6}A), demonstrating the robustness of the intrinsic property of our model to anticipation.

Although STPP shows promise as an underlying mechanism for sensory prediction, certain results of our model cannot yet be corroborated by current animal experiments. In our model, the anticipatory time varied with the angular velocities of the stimuli, which usually rose first and then fell. However, the average anticipatory time measured in a population of HD cells in the anterior thalamus tended to be approximately constant over a broad range of head turn velocities \citep{Goodridge2000}. Differently, some studies reported other thought-provoking issues. The anticipatory time varied among different HD cells in the anterior thalamus \citep{Blair1997}, and the complex results obscured the dependency of the anticipation time on the angular velocity \citep{Bassett2005}. Briefly, due to the considerable variability in real data obtained from the HD cell system, it is challenging to determine the stability of the anticipatory time. More notably, if our proposed mechanism is correct, the differing STPP strength of cells would explain the great complexity of the dependence of the anticipatory time.

In conclusion, our work offers a novel feedforward framework that potentially encodes neural information processing based on STPP in the HD cell system. A consideration of heterogeneous interactions in the neural system shows that NMDAR-dependent STPP may coordinate with other mechanisms, an understanding of which would provide a more comprehensive account of neural information processing. An improved understanding of brain processing and networking may also inspire more computational algorithms for prediction. In the future, assessing the effects of our proposed feedforward framework with other sensory inputs, such as sounds with increasing frequencies and predictable spatial locations, will be a crucial step in validating its applicability.


\section*{Author Contributions}
H.Z., S.C.Y, and C.C.A.F. conceptualized and designed the study; H.Z. made the computational modeling; H.Z. and C.C.A.F performed the data analysis; H.Z., S.C.Y, and C.C.A.F.  wrote the paper; H.Z., S.C.Y, and C.C.A.F.  revised the paper; S.C.Y and C.C.A.F acquired funding.

\section*{Funding}
Grants from the Research Grants Council of Hong Kong (11102120 and 11101922) for S.C.Y and startup grant (9610591) from City University of Hong Kong to C.C.A.F.

\section*{Acknowledgments}
The Figure \ref{fig:fig1}B was created with \href{https://www.biorender.com/}{BioRender.com}.

\section*{Supplementary Material}
The Supplementary Material for this article can be found online at: (to be confirmed).

\section*{Source Code Availability Statement}
{ The source codes used in this study are available at: \href{https://github.com/fccaa/cann_stpp_2023}{https://github.com/fccaa/cann\_stpp\_2023}.}

\bibliographystyle{Frontiers-Harvard} 
\bibliography{the_bib}


\section*{Figure captions}


\clearpage
\newpage 
\setcounter{section}{0}
\setcounter{figure}{0}
\renewcommand{\thefigure}{S\arabic{figure}}
\section*{Supplementary Material}

\section{Translational Stability of the Dynamical System}
In order to study the translational stability issue of static solutions of CANN with STPP. We first looked for the stationary states by considering temporal derivatives in Equations (1), (5) and (6) to be zero and substituted the results into Equations (2) and (4), then we had
\begin{align}
u_{0}\left(x\right) & =\left[1+S_{0}(x)\right]I_0^{\text{tot}}\left(x\right),\\
S_{0}\left(x\right) & =\tau_{1}\alpha Q_{0}\left(x\right)f_{S}\left[r_{0}\left(x\right)\right],\\
Q_{0}\left(x\right) & =\tau_{2}\beta\left[1-Q_{0}\left(x\right)\right]f_{Q}\left[I_{0}^{\text{tot}}\left(x\right)\right]-\tau_{2}\alpha Q_{0}\left(x\right)f_{S}\left[r_{0}\left(x\right)\right],\\
I_{0}^{\text{tot}}\left(x\right) & =\int dx^{\prime}J\left(x,x^{\prime}\right)r_{0}\left(x^{\prime}\right),\\
r_0\left(x\right)&=
\frac{u_{0}\left(x\right)^{2}}{1+\frac{1}{8\sqrt{2\pi}a}k\int dx^{\prime}u_{0}\left(x^{\prime}\right)^{2}}.
\end{align}
Since the functional forms were complicated, the stationary states were solved by numerical methods through simulations.

Next, we considered the network states with a small positional displacement to be
\begin{align}
u\left(x,t\right) & =u_{0}\left(x\right)+u_{1}\left(t\right)\frac{du_{0}\left(x\right)}{dx},\\
S\left(x,t\right) & =S_{0}\left(x\right)+S_{1}\left(t\right)\left(x-z\right)S_{0}\left(x\right),\\
Q\left(x,t\right) & =Q_{0}\left(x\right)+Q_{1}\left(t\right)\left(x-z\right)Q_{0}\left(x\right),
\end{align}
where $u_{0}\left(x\right)$, $S_{0}\left(x\right)$, and $Q_{0}\left(x\right)$ and $u_{1}\left(t\right)$, $S_{1}\left(t\right)$, and $Q_{1}\left(t\right)$ are the stationary states and the displacements of $u\left(x,t\right)$, $S\left(x,t\right)$, and $Q\left(x,t\right)$, respectively. $z$ is the center of mass of $u_{0}\left(x\right)$.

Then the function $r\left(x, t\right)$ became
\begin{align}
r\left(x,t\right) & =\frac{u_{0}\left(x\right)^{2}+2u_{0}\left(x\right)u_{1}\left(t\right)\frac{du_{0}\left(x\right)}{dx}}{1+\frac{1}{8\sqrt{2\pi}a}k\int dx^{\prime}\left[u_{0}\left(x^{\prime}\right)^{2}+2u_{0}\left(x^{\prime}\right)u_{1}\left(t\right)\frac{du_{0}\left(x^{\prime}\right)}{dx}\right]}\nonumber \\
 & =\frac{u_{0}\left(x\right)^{2}+2u_{0}\left(x\right)u_{1}\left(t\right)\frac{du_{0}\left(x\right)}{dx}}{1+\frac{1}{8\sqrt{2\pi}a}k\int dx^{\prime}u_{0}\left(x^{\prime}\right)^{2}}\nonumber \\
 & =r_{0}\left(x\right)+\frac{2u_{0}\left(x\right)u_{1}\left(t\right)}{B}\frac{du_{0}\left(x\right)}{dx},\\
B&=1+\frac{1}{8\sqrt{2\pi}a}k\int dx^{\prime}u_{0}\left(x^{\prime}\right)^{2}.
\end{align}
The function $I^{\text{tot}}\left(x,t\right)$ became
\begin{align}
I^{\text{tot}}\left(x,t\right) & =\int dx^{\prime}J\left(x,x^{\prime}\right)r\left(x^{\prime},t\right)\nonumber\\
 & =\int dx^{\prime}J\left(x,x^{\prime}\right)\left[r_{0}\left(x\right)+\frac{2u_{0}\left(x\right)u_{1}\left(t\right)}{B}\frac{du_{0}\left(x\right)}{dx}\right]\nonumber\\
 & =\int dx^{\prime}J\left(x,x^{\prime}\right)r_{0}\left(x^{\prime}\right)+\frac{2u_{1}\left(t\right)}{B}\int dx^{\prime}J\left(x,x^{\prime}\right)u_{0}\left(x^{\prime}\right)\frac{du_{0}\left(x^{\prime}\right)}{dx}\nonumber\\
 & =I_{0}^{\text{tot}}\left(x\right)+\frac{2u_{1}\left(t\right)}{B}\int dx^{\prime}J\left(x,x^{\prime}\right)u_{0}\left(x^{\prime}\right)\frac{du_{0}\left(x^{\prime}\right)}{dx}.
\end{align}

For Equation (1), we had
\begin{align}
\tau_\text{s}\frac{du_{1}\left(t\right)}{dt}\frac{du_{0}\left(x\right)}{dx} & =-u_{0}\left(x\right)-u_{1}\left(t\right)\frac{du_{0}\left(x\right)}{dx}+\left[1+S_{0}\left(x\right)+S_{1}\left(t\right)\left(x-z\right)S_{0}\left(x\right)\right]\nonumber\\
 & \ \ \ \ \ \ \ \ \times\int dx^{\prime}J\left(x,x^{\prime}\right)\left[r_{0}\left(x^{\prime}\right)+\frac{2u_{0}\left(x^{\prime}\right)u_{1}\left(t\right)}{B}\frac{du_{0}\left(x^{\prime}\right)}{dx}\right]\nonumber \\
 & =-\cancel{u_{0}\left(x\right)}-u_{1}\left(t\right)\frac{du_{0}\left(x\right)}{dx}+\cancel{\left[1+S_{0}\left(x\right)\right]\int dx^{\prime}J\left(x,x^{\prime}\right)r_{0}\left(x\right)}\nonumber \\
 & \ \ \ \ \ \ \ \ +2\frac{u_{1}\left(t\right)}{B}\left[1+S_{0}\left(x\right)\right]\int dx^{\prime}J\left(x,x^{\prime}\right)u_{0}\left(x^{\prime}\right)\frac{du_{0}\left(x^{\prime}\right)}{dx}\nonumber \\
 & \ \ \ \ +S_{1}\left(t\right)\left(x-z\right)S_{0}\left(x\right)\int dx^{\prime}J\left(x,x^{\prime}\right)r_{0}\left(x^{\prime}\right)\nonumber \\
 & \ \ \ \ \ \ \ \ +2\frac{u_{1}\left(t\right)}{B}S_{1}\left(t\right)\left(x-z\right)S_{0}\left(x\right)\int dx^{\prime}J\left(x,x^{\prime}\right)u_{0}\left(x^{\prime}\right)\frac{du_{0}\left(x^{\prime}\right)}{dx}.
\end{align}
By gathering odd terms, we had
\begin{align}
\tau_\text{s}\frac{du_{1}\left(t\right)}{dt}\frac{du_{0}\left(x\right)}{dx} & =-u_{1}\left(t\right)\frac{du_{0}\left(x\right)}{dx}+2\frac{u_{1}\left(t\right)}{B}\left[1+S_{0}\left(x\right)\right]\int dx^{\prime}J\left(x,x^{\prime}\right)u_{0}\left(x^{\prime}\right)\frac{du_{0}\left(x^{\prime}\right)}{dx}\nonumber \\
 & \ \ \ \ +S_{1}\left(t\right)\left(x-z\right)S_{0}\left(x\right)\int dx^{\prime}J\left(x,x^{\prime}\right)r_{0}\left(x^{\prime}\right),\\
\frac{du_{1}\left(t\right)}{dt} & =-u_{1}\left(t\right)\frac{1}{\tau_\text{s}}\Biggl\{1-\frac{2}{B}\times\frac{1}{\int dx^{\prime}\left[\frac{du_{0}\left(x^{\prime}\right)}{dx}\right]^{2}}\nonumber \\
 & \ \ \ \ \ \ \ \ \ \ \ \ \times\int dx\frac{du_{0}\left(x\right)}{dx}\left[1+S_{0}\left(x\right)\right]\int dx^{\prime}J\left(x,x^{\prime}\right)u_{0}\left(x^{\prime}\right)\frac{du_{0}\left(x^{\prime}\right)}{dx}\Biggr\}\nonumber \\
 & \ \ \ \ +S_{1}\left(t\right)\frac{1}{\tau_\text{s}}\frac{1}{\int dx^{\prime}\left[\frac{du_{0}\left(x^{\prime}\right)}{dx}\right]^{2}}\int dx\frac{du_{0}\left(x\right)}{dx}\left(x-z\right)S_{0}\left(x\right)\int dx^{\prime}J\left(x,x^{\prime}\right)r_{0}\left(x^{\prime}\right).
\end{align}
For Equation (5), we had
\begin{align}
\frac{dS_{1}\left(t\right)}{dt}\left(x-z\right)S_{0}\left(x\right) & =-\frac{1}{\tau_{1}}S_{0}\left(x\right)-\frac{1}{\tau_{1}}S_{1}\left(t\right)\left(x-z\right)S_{0}\left(x\right)+\alpha\left[Q_{0}\left(x\right)+Q_{1}\left(t\right)\left(x-z\right)Q_{0}\left(x\right)\right]\nonumber \\
 & \ \ \ \ \ \ \ \ \ \ \ \times f_{S}\left[r_{0}\left(x\right)+\frac{2u_{0}\left(x\right)u_{1}\left(t\right)}{B}\frac{du_{0}\left(x\right)}{dx}\right]\nonumber \\
& =-\frac{1}{\tau_{1}}S_{0}\left(x\right)-\frac{1}{\tau_{1}}S_{1}\left(t\right)\left(x-z\right)S_{0}\left(x\right)\nonumber \\
 & \ \ \ \ +\alpha\left[Q_{0}\left(x\right)+Q_{1}\left(t\right)\left(x-z\right)Q_{0}\left(x\right)\right]\nonumber \\
 & \ \ \ \ \ \ \ \ \times\left\{ f_{S}\left[r_{0}\left(x\right)\right]+f_{S}^{\prime}\left[r_{0}\left(x\right)\right]\frac{2u_{0}\left(x\right)u_{1}\left(t\right)}{B}\frac{du_{0}\left(x\right)}{dx}\right\} \nonumber \\
 & =\cancel{-\frac{1}{\tau_{1}}S_{0}\left(x\right)+\alpha Q_{0}\left(x\right)f_{S}\left[r_{0}\left(x\right)\right]}\nonumber \\
 & \ \ \ \ -\frac{1}{\tau_{1}}S_{1}\left(t\right)\left(x-z\right)S_{0}\left(x\right)+\alpha Q_{1}\left(t\right)\left(x-z\right)Q_{0}\left(x\right)f_{S}\left[r_{0}\left(x\right)\right]\nonumber \\
 & \ \ \ \ \ \ \ \ \ +\alpha Q_{0}\left(x\right)f_{S}^{\prime}\left[r_{0}\left(x\right)\right]\frac{2u_{0}\left(x\right)u_{1}\left(t\right)}{B}\frac{du_{0}\left(x\right)}{dx}\nonumber \\
 & \ \ \ \ +\alpha Q_{1}\left(t\right)\left(x-z\right)Q_{0}\left(x\right)f_{S}^{\prime}\left[r_{0}\left(x\right)\right]\frac{2u_{0}\left(x\right)u_{1}\left(t\right)}{B}\frac{du_{0}\left(x\right)}{dx}.
\end{align}
By gathering odd terms, we had
\begin{align}
\frac{dS_{1}\left(t\right)}{dt}\left(x-z\right)S_{0}\left(x\right) & = -\frac{1}{\tau_{1}}S_{1}\left(t\right)\left(x-z\right)S_{0}\left(x\right)+\alpha Q_{1}\left(t\right)\left(x-z\right)Q_{0}\left(x\right)f_{S}\left[r_{0}\left(x\right)\right]\nonumber \\
 & \ \ \ \ +\alpha Q_{0}\left(x\right)f_{S}^{\prime}\left[r_{0}\left(x\right)\right]\frac{2u_{0}\left(x\right)u_{1}\left(t\right)}{B}\frac{du_{0}\left(x\right)}{dx},\\
\frac{dS_{1}\left(t\right)}{dt} & =-\frac{1}{\tau_{1}}S_{1}\left(t\right)+Q_{1}\left(t\right)\alpha\frac{1}{\int dx^{\prime}\left[\left(x^{\prime}-z\right)S_{0}\left(x^{\prime}\right)\right]^{2}}\nonumber \\
& \ \ \ \ \ \ \ \ \times\int dx\left(x-z\right)S_{0}\left(x\right)\left(x-z\right)Q_{0}\left(x\right)f_{S}\left[r_{0}\left(x\right)\right]\nonumber \\
 & \ \ \ \ +u_{1}\left(t\right)\frac{2\alpha}{B}\frac{1}{\int dx^{\prime}\left[\left(x^{\prime}-z\right)S_{0}\left(x^{\prime}\right)\right]^{2}}\nonumber \\
 & \ \ \ \ \ \ \ \ \times\int dx\left(x-z\right)S_{0}\left(x\right)Q_{0}\left(x\right)f_{S}^{\prime}\left[r_{0}\left(x\right)\right]u_{0}\left(x\right)\frac{du_{0}\left(x\right)}{dx}.
\end{align}
For Equation (6), we had
\begin{align}
\frac{dQ_{1}\left(t\right)}{dt}\left(x-z\right)Q_{0}\left(x\right) 
& =-\frac{Q_{0}\left(x\right)+Q_{1}\left(t\right)\left(x-z\right)Q_{0}\left(x\right)}{\tau_{2}}\nonumber \\
& \ \ \ \ -\alpha\left[Q_{0}\left(x\right)+Q_{1}\left(t\right)\left(x-z\right)Q_{0}\left(x\right)\right]f_{S}\left[r\left(x,t\right)\right]\nonumber \\
& \ \ \ \ \ \ \ \ +\beta\left[1-Q_{0}\left(x\right)-Q_{1}\left(t\right)\left(x-z\right)Q_{0}\left(x\right)\right]f_{Q}\left[I^{\text{tot}}\left(x,t\right)\right]\nonumber \\
& =-\frac{Q_{0}\left(x\right)}{\tau_{2}}-\alpha Q_{0}\left(x\right)f_{S}\left[r\left(x,t\right)\right]+\beta\left[1-Q_{0}\left(x\right)\right]f_{Q}\left[I^{\text{tot}}\left(x,t\right)\right]\nonumber \\
& \ \ \ \ -\frac{Q_{1}\left(t\right)}{\tau_{2}}\left(x-z\right)Q_{0}\left(x\right)-\alpha Q_{1}\left(t\right)\left(x-z\right)Q_{0}\left(x\right)f_{S}\left[r\left(x,t\right)\right]\nonumber \\
& \ \ \ \ \ \ \ \ -\beta Q_{1}\left(t\right)\left(x-z\right)Q_{0}\left(x\right)f_{Q}\left[I^{\text{tot}}\left(x,t\right)\right]\nonumber \\
& =\cancel{-\frac{Q_{0}\left(x\right)}{\tau_{2}}-\alpha Q_{0}\left(x\right)f_{S}\left[r_{0}\left(x\right)\right]}-\alpha u_{1}\left(t\right)\frac{2}{B}Q_{0}\left(x\right)f_{S}^{\prime}\left[r_{0}\left(x\right)\right]u_{0}\left(x\right)\frac{du_{0}\left(x\right)}{dx}\nonumber \\
& \ \ \ \ \cancel{+\beta\left[1-Q_{0}\left(x\right)\right]f_{Q}\left[I_{0}^{\text{tot}}\left(x\right)\right]}\nonumber \\
& \ \ \ \ \ \ \ \ +\beta\frac{2}{B}u_{1}\left(t\right)\left[1-Q_{0}\left(x\right)\right]f_{Q}^{\prime}\left[I_{0}^{\text{tot}}\left(x\right)\right]\int dx^{\prime}J\left(x,x^{\prime}\right)u_{0}\left(x^{\prime}\right)\frac{du_{0}\left(x^{\prime}\right)}{dx}\nonumber \\
& \ \ \ \ -\frac{Q_{1}\left(t\right)}{\tau_{2}}\left(x-z\right)Q_{0}\left(x\right)-\alpha Q_{1}\left(t\right)\left(x-z\right)Q_{0}\left(x\right)f_{S}\left[r_{0}\left(x\right)\right]\nonumber \\
& \ \ \ \ \ \ \ \ -\alpha\frac{2}{B} Q_{1}\left(t\right)u_{1}\left(t\right)\left(x-z\right)Q_{0}\left(x\right)f_{S}^{\prime}\left[r_{0}\left(x\right)\right]u_{0}\left(x\right)\frac{du_{0}\left(x\right)}{dx}\nonumber \\
& \ \ \ \ -\beta Q_{1}\left(t\right)\left(x-z\right)Q_{0}\left(x\right)f_{Q}\left[I_{0}^{\text{tot}}\left(x\right)\right]\nonumber \\
& \ \ \ \ \ \ \ \ -\beta\frac{2}{B}Q_{1}\left(t\right)u_{1}\left(t\right)\left(x-z\right)Q_{0}\left(x\right)f_{Q}^{\prime}\left[I_{0}^{\text{tot}}\left(x\right)\right]\nonumber \\
& \ \ \ \ \ \ \ \ \ \ \ \ \times\int dx^{\prime}J\left(x,x^{\prime}\right)u_{0}\left(x^{\prime}\right)\frac{du_{0}\left(x^{\prime}\right)}{dx}.
\end{align}
By gathering odd terms, we had
\begin{align}
\frac{dQ_{1}\left(t\right)}{dt}\left(x-z\right)Q_{0}\left(x\right) & =-u_{1}\left(t\right)\times\Biggl\{\alpha\frac{2}{B}Q_{0}\left(x\right)f_{S}^{\prime}\left[r_{0}\left(x\right)\right]u_{0}\left(x\right)\frac{du_{0}\left(x\right)}{dx}\nonumber \\
 & \ \ \ \ -\beta\frac{2}{B}\left[1-Q_{0}\left(x\right)\right]f_{Q}^{\prime}\left[I_{0}^{\text{tot}}\left(x\right)\right]\int dx^{\prime}J\left(x,x^{\prime}\right)u_{0}\left(x^{\prime}\right)\frac{du_{0}\left(x^{\prime}\right)}{dx}\Biggr\}\nonumber \\
 & \ \ \ \ \ \ \ \ -Q_{1}\left(t\right)\times\Biggl\{\frac{1}{\tau_{2}}\left(x-z\right)Q_{0}\left(x\right)+\alpha\left(x-z\right)Q_{0}\left(x\right)f_{S}\left[r_{0}\left(x\right)\right]\nonumber \\
 & \ \ \ \ +\beta\left(x-z\right)Q_{0}\left(x\right)f_{Q}\left[I_{0}^{\text{tot}}\left(x\right)\right]\Biggr\},\\
\frac{dQ_{1}\left(t\right)}{dt} & =-u_{1}\left(t\right)\times\frac{1}{\int dx^{\prime}\left[\left(x^{\prime}-z\right)Q_{0}\left(x^{\prime}\right)\right]^{2}}\int dx\left(x-z\right)Q_{0}\left(x\right)\nonumber \\
 & \ \ \ \ \ \ \ \ \times\Biggl\{\alpha\frac{2}{B}Q_{0}\left(x\right)f_{S}^{\prime}\left[r_{0}\left(x\right)\right]u_{0}\left(x\right)\frac{du_{0}\left(x\right)}{dx}\nonumber \\
 & \ \ \ \ \ \ \ \ \ \ \ \ -\beta\frac{2}{B}\left[1-Q_{0}\left(x\right)\right]f_{Q}^{\prime}\left[I_{0}^{\text{tot}}\left(x\right)\right]\int dx^{\prime}J\left(x,x^{\prime}\right)u_{0}\left(x^{\prime}\right)\frac{du_{0}\left(x^{\prime}\right)}{dx}\Biggr\}\nonumber \\
 & \ \ \ \ -Q_{1}\left(t\right)\times\frac{1}{\int dx^{\prime}\left[\left(x^{\prime}-z\right)Q_{0}\left(x^{\prime}\right)\right]^{2}}\int dx\left(x-z\right)Q_{0}\left(x\right)\nonumber \\
 & \ \ \ \ \ \ \ \ \times\Biggl\{\frac{1}{\tau_{2}}\left(x-z\right)Q_{0}\left(x\right)+\alpha\left(x-z\right)Q_{0}\left(x\right)f_{S}\left[r_{0}\left(x\right)\right]\nonumber \\
& \ \ \ \ \ \ \ \ \ \ \ \ +\beta\left(x-z\right)Q_{0}\left(x\right)f_{Q}\left[I_{0}^{\text{tot}}\left(x\right)\right]\Biggr\}.
\end{align}

In summary, we had 
\begin{equation}
\frac{d}{dt}\left(\begin{array}{c}
u_{1}\left(t\right)\\
S_{1}\left(t\right)\\
Q_{1}\left(t\right)
\end{array}\right)=\left(\begin{array}{ccc}
M_{uu} & M_{uS} & 0\\
M_{Su} & M_{SS} & M_{SQ}\\
M_{Qu} & 0 & M_{QQ}
\end{array}\right)\left(\begin{array}{c}
u_{1}\left(t\right)\\
S_{1}\left(t\right)\\
Q_{1}\left(t\right)
\end{array}\right),
\end{equation}
where
\begin{align}
M_{uu} & =-\frac{1}{\tau_\text{s}}\Biggl\{1-\frac{2}{B}\times\frac{1}{\int dx^{\prime}\left[\frac{du_{0}\left(x^{\prime}\right)}{dx}\right]^{2}}\nonumber\\
& \ \ \ \ \ \ \ \ \times\int dx\frac{du_{0}\left(x\right)}{dx}\left[1+S_{0}\left(x\right)\right]\int dx^{\prime}J\left(x,x^{\prime}\right)u_{0}\left(x^{\prime}\right)\frac{du_{0}\left(x^{\prime}\right)}{dx}\Biggr\},\\
M_{uS} & =\frac{1}{\tau_\text{s}}\frac{1}{\int dx^{\prime}\left[\frac{du_{0}\left(x^{\prime}\right)}{dx}\right]^{2}}\int dx\frac{du_{0}\left(x\right)}{dx}\left(x-z\right)S_{0}\left(x\right)I_{0}^{\text{tot}}\left(x\right),\\
M_{Su} & =\frac{2\alpha}{B}\frac{1}{\int dx^{\prime}\left[\left(x^{\prime}-z\right)S_{0}\left(x^{\prime}\right)\right]^{2}}\int dx\left(x-z\right)S_{0}\left(x\right)Q_{0}\left(x\right)f_{S}^{\prime}\left[r_{0}\left(x\right)\right]u_{0}\left(x\right)\frac{du_{0}\left(x\right)}{dx},\\
M_{SS} & =-\frac{1}{\tau_{1}},\\
M_{SQ} & =\alpha\frac{1}{\int dx^{\prime}\left[\left(x^{\prime}-z\right)S_{0}\left(x^{\prime}\right)\right]^{2}}\int dx\left(x-z\right)S_{0}\left(x\right)\left(x-z\right)Q_{0}\left(x\right)f_{S}\left[r_{0}\left(x\right)\right],\\
M_{Qu} & =-\frac{1}{\int dx^{\prime}\left[\left(x^{\prime}-z\right)Q_{0}\left(x^{\prime}\right)\right]^{2}}\int dx\left(x-z\right)Q_{0}\left(x\right)\Biggl\{\alpha\frac{2}{B}Q_{0}\left(x\right)f_{S}^{\prime}\left[r_{0}\left(x\right)\right]u_{0}\left(x\right)\frac{du_{0}\left(x\right)}{dx}\nonumber \\
 & \ \ \ \ \ \ \ \ -\beta\frac{2}{B}\left[1-Q_{0}\left(x\right)\right]f_{Q}^{\prime}\left[I_{0}^{\text{tot}}\left(x\right)\right]\int dx^{\prime}J\left(x,x^{\prime}\right)u_{0}\left(x^{\prime}\right)\frac{du_{0}\left(x^{\prime}\right)}{dx}\Biggr\},\\
 M_{QQ} & =-\frac{1}{\tau_{2}}-\frac{1}{\int dx^{\prime}\left[\left(x^{\prime}-z\right)Q_{0}\left(x^{\prime}\right)\right]^{2}}\int dx\left(x-z\right)Q_{0}\left(x\right)\nonumber \\
 & \ \ \ \ \ \ \ \ \times\Biggl\{\alpha\left(x-z\right)Q_{0}\left(x\right)f_{S}\left[r_{0}\left(x\right)\right]+\beta\left(x-z\right)Q_{0}\left(x\right)f_{Q}\left[I_{0}^{\text{tot}}\left(x\right)\right]\Biggr\}.
\end{align}



\section{Supplementary Figure}






\begin{figure}[htbp]
\begin{center}
\includegraphics[width=16cm]{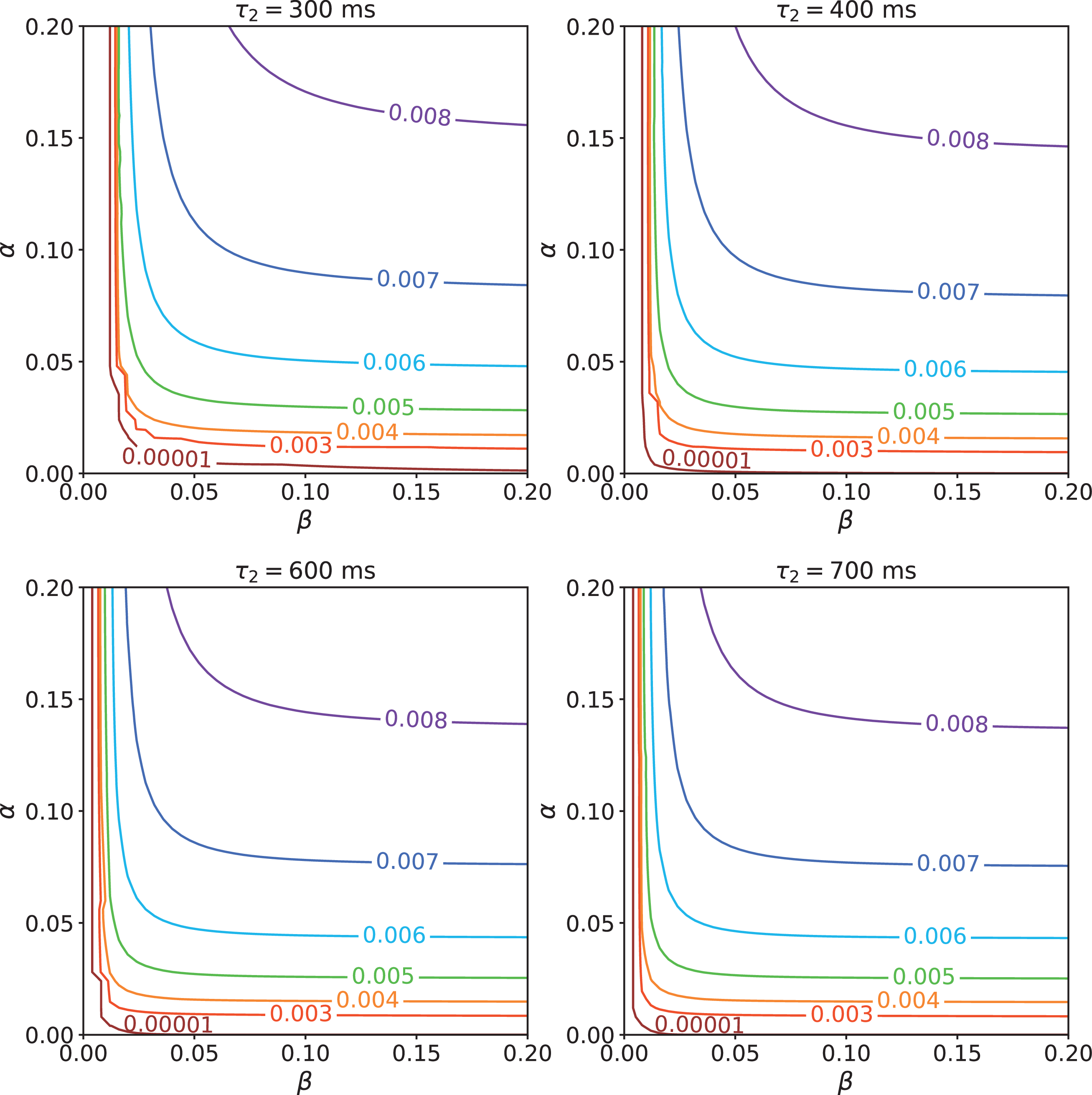}\label{fig:figs1}
\end{center}
\caption{Contour maps of $v_{\text{int}}$ when $\tau_2=300$ ms, $\tau_2=400$ ms, $\tau_2=600$ ms, and $\tau_2=700$ ms. The intrinsic speeds under these conditions share similar patterns and ranges with those obtained when $\tau_2=500$ ms (Figure 6A), indicating the robustness of the intrinsic property of the model to anticipation. The unit of $v_{\text{int}}$ is rad/ms. Parameters: $a=0.5$, $k=0.5$.}\label{fig:1}
\end{figure}





\end{document}